\documentclass[11pt]{article}

\usepackage[final]{acl}

\usepackage{times}
\usepackage{latexsym}

\usepackage[T1]{fontenc}

\usepackage[utf8]{inputenc}

\usepackage{microtype}

\usepackage{inconsolata}

\usepackage{graphicx}

\usepackage[most]{tcolorbox}
\usepackage{xcolor}

\definecolor{myorange}{RGB}{242,153,74}
\definecolor{lightbox}{RGB}{245,232,220}
\usepackage{tikz}
\usetikzlibrary{arrows.meta, shapes, positioning, calc, backgrounds}

\usepackage[edges]{forest}

\usepackage{graphicx}
\usepackage{amsmath}
\usepackage{amssymb}
\usepackage{booktabs}
\usepackage{caption}
\usepackage{multirow}
\usepackage{xurl}
\usepackage{booktabs}
\usepackage{graphicx}
\usepackage{colortbl}

\usepackage{threeparttable}
\usepackage{hyperref}

\title{Resource Consumption Threats in Large Language Models}

\author{
 \textbf{Yuanhe Zhang\textsuperscript{1}}, 
 \textbf{Xinyue Wang\textsuperscript{1}}, 
 \textbf{Zhican Chen\textsuperscript{1}}, 
 \textbf{Weiliu Wang\textsuperscript{2}}, 
 \textbf{Zilu Zhang\textsuperscript{1}}, 
 \\
 \textbf{Zhengshuo Gong\textsuperscript{1}},
 \textbf{Zhenhong Zhou\textsuperscript{3}},
 \textbf{Kun Wang\textsuperscript{3}},
 \textbf{Li Sun\textsuperscript{1}},
 \textbf{Yang Liu\textsuperscript{3}},
 \textbf{Sen Su\textsuperscript{4, 1,  $^\dagger$}} 
\\ \textsuperscript{\rm 1}Beijing University of Posts and Telecommunications \quad
\textsuperscript{\rm 2}Hangzhou Dianzi University
\\ \textsuperscript{\rm 3}Nanyang Technological University \quad
\textsuperscript{\rm 4}Chongqing University of Posts and Telecommunications
\\ \{charmes-zhang, susen\}@bupt.edu.cn;
}

\begin{document}
\maketitle

\begingroup
\renewcommand\thefootnote{}\footnotemark
\footnotetext{$\dagger$ indicates corresponding author.}
\endgroup

\begin{abstract}

Given limited and costly computational infrastructure, resource efficiency is a key requirement for large language models (LLMs). 
Efficient LLMs increase service capacity for providers and reduce latency and API costs for users.
Recent resource consumption threats induce excessive generation, degrading model efficiency and harming both service availability and economic sustainability.
This survey presents a systematic review of threats to resource consumption in LLMs. 
We further establish a unified view of this emerging area by clarifying its scope and examining the problem along the full pipeline from threat induction to mechanism understanding and mitigation. Our goal is to clarify the problem landscape for this emerging area, thereby providing a clearer foundation for characterization and mitigation.
Our data is open source on~\url{https://github.com/shuita2333/resource-consumption-threats-in-llms}.
\end{abstract}

\section{Introduction}
\label{sec:intro}

Large language models~\cite{vaswani2017attention} operate under limited and costly computational infrastructure~\cite{samsi2023words,miao2024spotserve}, making resource efficiency a core requirement for practical deployment. Efficient resource usage improves service throughput for providers and reduces API costs for users. Consequently, improving computational efficiency has been studied as an engineering problem~\cite{zhou2024survey,fernandez2025energy} across LLMs~\cite{vaswani2017attention}, reasoning large language models (RLLMs)~\cite{wei2022chain}, multimodal large language models (MLLMs)~\cite{alayrac2022flamingo}, and LLM-based agentic environments (Agents)~\cite{yao2022react}. 
However, optimization alone is insufficient to address this threat, as many adversarial resource consumption attacks still threaten the usability and sustainability of LLMs~\cite{zhang2025crabs}.

\begin{figure}[t]
    \centering \includegraphics[width=1\columnwidth]{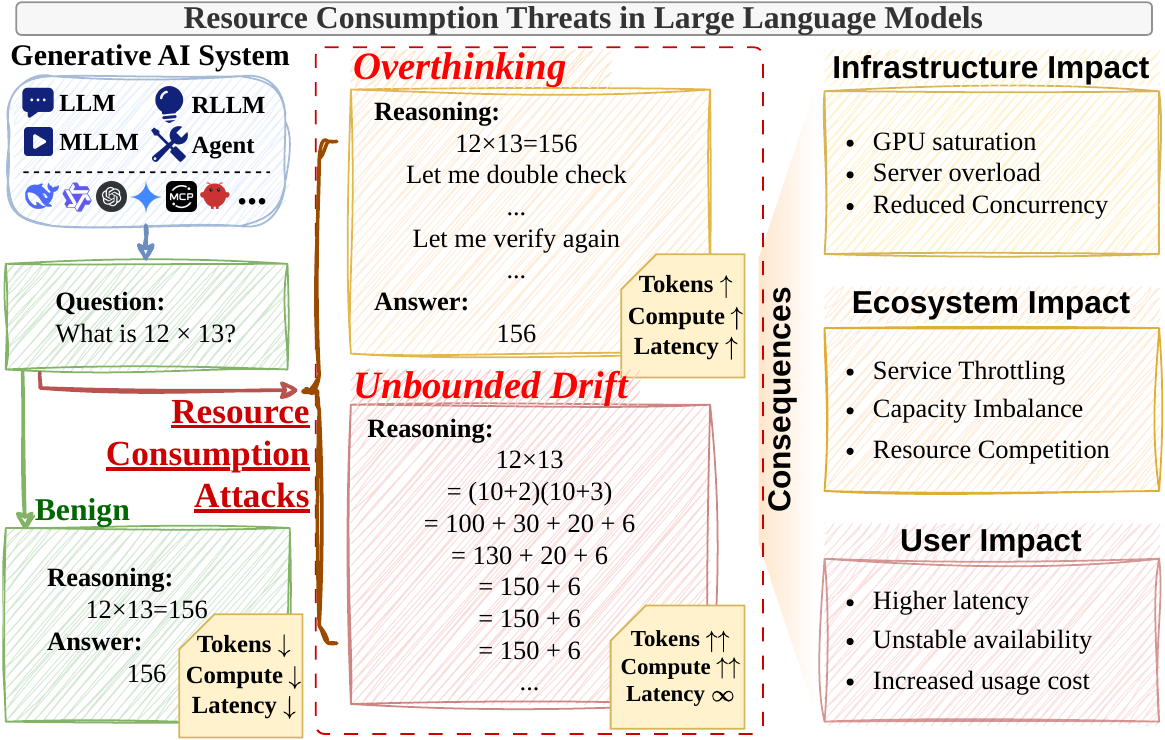}
    \caption{Overview of resource consumption threats in large language models.}
    \label{fig:attack}
    \vspace{-9pt}
\end{figure}

Resource consumption attacks are designed to induce disproportionate computational overhead in LLMs~\cite{shumailov2021sponge}. Recent studies suggest that such attacks have emerged as an important safety risk~\cite{geiping2024coercing, gao2024denial}.
As shown in Figure~\ref{fig:attack}, rather than directly inducing harmful outputs or extracting private information, they trigger excessive and unnecessary generation~\cite{li2025loopllm}, which degrades model throughput, inflates operational costs, and places excessive pressure on shared service resources~\cite{gao2025resource}.
This shift reframes resource usage in LLMs from a performance issue to a core safety concern.
Although research on this problem has begun to appear across diverse model settings~\cite{wang2023energy}, the field remains fragmented by inconsistent terminology and threat assumptions. As a result, the scope and core research questions of resource consumption attacks remain insufficiently clarified.

\begin{figure*}[t]
    \centering \includegraphics[width=1\textwidth]{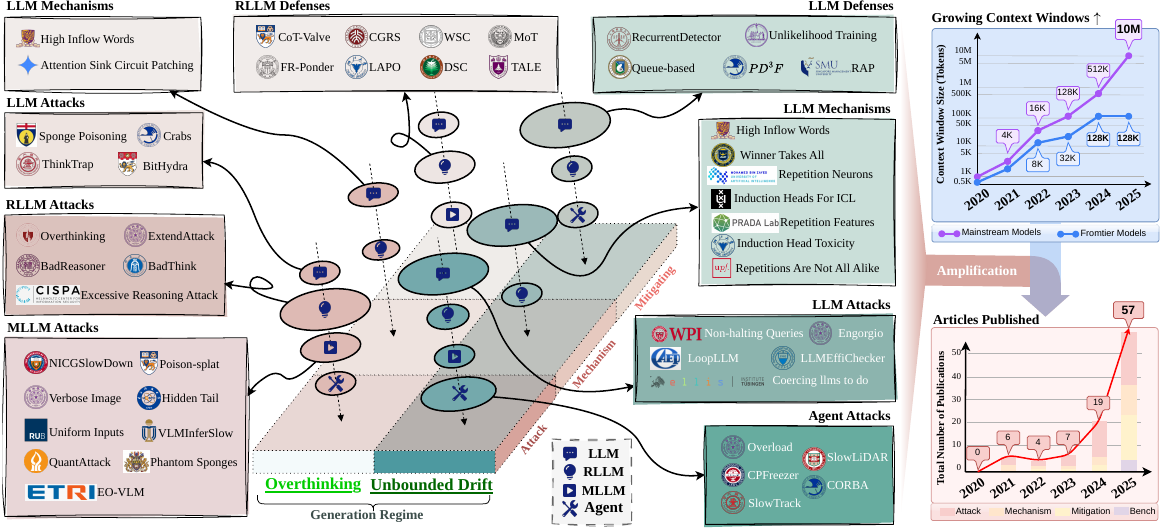}
    \caption{A unified view of resource consumption threats in large language models.}
    \label{fig:main}
    \vspace{-9pt}
\end{figure*}

Our review centers on \textit{resource consumption threats in large language models}, especially malicious behaviors that amplify computational costs through extended or uncontrolled generation.
To provide a unified view, we organize this landscape using an effect-oriented taxonomy with two representative regimes: \textbf{Overthinking}, where generation remains relevant to the task but exceeds practical utility, and \textbf{Unbounded Drift}, where the generation trajectory becomes progressively less controlled and less convergent. 
We also discuss selected studies from machine learning and early deep learning architectures as related evidence that helps contextualize generative resource abuse.
As shown in Figure~\ref{fig:main}, this taxonomy clarifies the scope of resource consumption threats and links attacks to mechanism analysis and mitigation.

Based on this perspective, this survey makes three main contributions. First, we provide a comprehensive overview of recent advances on resource consumption threats.
Second, we introduce a unified taxonomy that organizes resource consumption risks into two representative regimes, Overthinking and Unbounded Drift, clarifying the scope of this safety problem. 
Third, we analyze open challenges in this emerging area and outline promising directions for future research.

\section{Safety Implications and Problem Taxonomy}
\label{sec:taxonomy}
\label{sec:classification}

\subsection{Attack Impact and System Risks}

\begin{figure}[t]
    \centering \includegraphics[width=1\columnwidth]{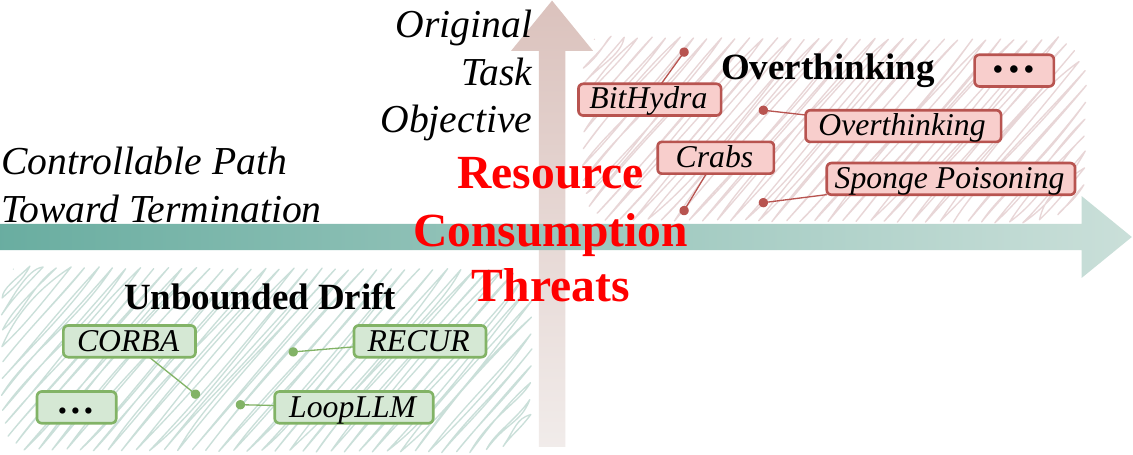}
    \caption{Taxonomy of resource consumption threats.}
    \label{fig:attack_taxonomy}
    \vspace{-9pt}
\end{figure}

Resource consumption attacks exploit LLMs' computational characteristics to induce excessive generation,  thereby degrading system effectiveness. The risks of such amplification are evident even in non-adversarial deployment settings. For example, the launch of DeepSeek-R1 attracted massive traffic, while the computational cost of its chain-of-thought reasoning was underestimated, together saturating available compute capacity and leading to service disruptions\footnote{\url{https://www.binance.com/en/square/post/01-26-2025-deepseek-r1-api-19448332337730}}. This example illustrates provider-side risks such as reduced throughput and increased operational cost. More broadly, such pressure can also propagate to users in the form of higher latency, unstable availability, and increased usage cost. 
Resource consumption, therefore, constitutes a systemic safety risk affecting both providers and users in the LLM ecosystem.
To better understand this emerging threat landscape, a clearer taxonomy is needed, as illustrated in Figure~\ref{fig:attack_taxonomy}.

\subsection{A Taxonomy of Generation Regimes}

Accordingly, while the survey reviews attacks, mechanisms, and defenses, the taxonomy focuses on the induced generation behavior. This distinction allows different attack forms to be analyzed under a common lens when they prolong generation in similar ways.
We distinguish them along two criteria: whether the extended generation remains aligned with the original task objective, and whether the generation process remains on a controllable path toward termination.

\textbf{Overthinking} refers to excessive generation that remains task-aligned and preserves normal stopping behavior, but incurs unnecessary cost through verbosity, redundant reasoning, or over-elaborate description. Its continuation is mainly sustained by task-relevant but low-utility elaboration.

\textbf{Unbounded Drift} refers to excessive generation in which the decoding trajectory is no longer reliably governed by the original task or by normal convergence dynamics. It typically manifests as repetitive loops, recursive self-extension, or self-reinforcing interaction chains that weaken semantic control, disrupt timely termination, or both.

The distinction is therefore not based on output length alone. Some cases may evolve from overthinking into drift; in such cases, we classify them by the dominant mechanism sustaining continuation. We provide a more detailed discussion of these impacts in Appendix~\ref{sec:appen_threat}.

\section{Threat Landscape and Related Precursors}
\label{sec:attack}
The threat landscape includes both core attacks in generative settings and several adjacent precursor lines. The former directly exploits the generative process to amplify computation, whereas the latter arises in earlier vision or system pipelines and are included here only when they help contextualize similar resource-amplification mechanisms.
\subsection{Early Forms of Attacks}

The challenge of resource consumption attacks originated in early deep learning architectures. Specifically, Sloth \cite{hong2020panda} revealed that gradient-based optimization could coerce DNNs into the most computationally expensive inference paths. Building on this concept of computational asymmetry, the seminal work on this attack, Sponge Examples \cite{shumailov2021sponge}, demonstrated that targeted perturbations of neural activations could dramatically increase energy consumption by disrupting hardware-level execution. This principle was quickly extended to Neural Machine Translation (NMT), where TranSlowDown \cite{chen2021transslowdown} and NMTSloth \cite{chen2022nmtsloth} showed that minute perturbations delaying the End-of-Sentence (EOS) token could force sequences toward maximum length limits, significantly inflating inference costs.

As research progressed, the attack surface expanded from inference-time perturbations to training-time backdoors. 
Prior work showed that efficiency-targeted backdoors can manipulate model behavior by hijacking inference control flow, including early model integrity attacks~\cite{cina2021hammer}, later studies on dynamic routing~\cite{chen2023dark}, and other deep architectures~\cite{meftah2025energy}.
This was further refined by \textit{Sponge Poisoning} \cite{cina2025energy, lintelo2024skipsponge}, which embeds backdoors to disrupt hardware-level sparsity and skip-connection logic. 
These early paradigms also exposed the practical severity of resource-oriented attacks on edge hardware~\cite{wang2023energy, hasan2025sponge}. 

Prior work showed that computational overhead can lead to direct physical consequences, including battery drain and device-level Denial-of-Service (DoS) attacks. 
These attacks reveal that adversarial inputs can amplify seemingly small perturbations into disproportionately high computational costs.

\subsection{Attack Surfaces in Large Language Models}

\subsubsection{Risks of Overthinking}

The most direct way to induce resource consumption is to artificially prolong model outputs. Sponge Poisoning in LLM \cite{cina2025energy} introduces a training-time backdoor that induces sustained output elongation and resource exhaustion. In black-box settings, Crabs \cite{zhang2025crabs} shows that semantic expansion via tree-structured queries can also effectively increase output length.
Beyond direct output prolongation, ThinkTrap \cite{li2025thinktrap} maps discrete tokens into a continuous embedding space and performs optimization in a low-dimensional subspace to induce excessive generation under black-box access.

At a more fine-grained level, resource consumption can even be induced by directly manipulating model parameters. BitHydra \cite{yan2025bithydra} proposes a bit-flip inference cost attack that alters model weights at the hardware-relevant level to compromise efficiency.
RepetitionCurse~\cite{huang2025repetitioncursemeasuringunderstandingrouter} shows that imbalanced expert routing in MoE can induce analogous repetitive failures.

These studies suggest that resource consumption vulnerabilities span multiple levels of the stack, from high-level semantic elicitation to low-level parameter and hardware-oriented manipulation.

\subsubsection{Risks of Unbounded Drift}

Although existing methods use different induction strategies, they often converge on a shared unbounded drift effect: once termination is destabilized, the generation process can enter repetitive or effectively non-terminating loops. Fixed points \cite{hammouri2025non}, and attention sinks \cite{yona2025interpreting} exploit intrinsic decoding dynamics, making repetitive trajectories easier to sustain. 
Others operationalize this effect more directly through entropy-driven optimization. LoopLLM \cite{li2025loopllm} induces repetitive generation loops through entropy-based search, while GCG \cite{geiping2024coercing}, Engorgio \cite{dong2024engorgio}, and LLMEffiChecker \cite{feng2024llmeffichecker} interfere with termination by manipulating critical tokens. In these attacks, the model is prevented from halting normally and is pushed toward a loop.

\begin{tcolorbox}[
  colback=lightbox,
  colframe=myorange,
  colbacktitle=myorange,
  coltitle=white,
  fonttitle=\bfseries,
  title=Key Insights.,
  rounded corners,
  arc=3mm
]

Current research lacks a cross-generational perspective, typically focusing on parameter scaling within a single model family rather than analyzing how these issues evolve as model capabilities grow. Furthermore, the field is heavily dominated by white-box methods, making it difficult to study commercial black-box environments. This is particularly true for unbounded drift.

\end{tcolorbox}

\subsection{Attack Surfaces in Reasoning Large Language Models}
Reasoning large language models (RLLMs) are prone to inefficiently prolonged generation: even on simple tasks such as 2+3, they may generate excessively long, redundant, and repetitive reasoning processes \cite{chen2025do}, which creates significant opportunities for attacks.

\subsubsection{Risks of Overthinking}

Chain-of-thought reasoning exposes a distinctive attack surface, as attackers can deliberately induce excessive and unnecessary reasoning.
Existing attacks achieve this through diverse strategies, including backdoor triggers in BadReasoner \cite{yi2025badreasonerplantingtunableoverthinking} and BadThink \cite{liu2025badthinktriggeredoverthinkingattacks}, context-based slowdown attacks that exploit retrieval or search mechanisms \cite{kumar2025overthinkslowdownattacksreasoning,zhu2025extendattackattackingserverslrms}, and adversarial perturbations that interfere with normal termination \cite{si2025excessivereasoningattackreasoning}. 
Together, these studies show that overthinking in reasoning models is not just an inefficiency but an exploitable vulnerability that prolongs reasoning and increases computational cost.

\subsubsection{Risks of Unbounded Drift}

Unbounded drift poses more severe risks than ordinary overthinking because it can trap reasoning models in self-perpetuating generation loops. 
RECUR~\cite{wang2026recur} shows that chain-of-thought reasoning may enter repeated reflective cycles, suggesting that such loop-like degeneration may arise broadly across generative architectures.

\begin{tcolorbox}[
  colback=lightbox,
  colframe=myorange,
  colbacktitle=myorange,
  coltitle=white,
  fonttitle=\bfseries,
  title=Key Insights.,
  rounded corners,
  arc=3mm
]

Recent studies have provided a relatively comprehensive understanding of overthinking in reasoning models, while also showing that unbounded drift poses a severe threat to stable and reliable generation.

\end{tcolorbox}

\subsection{Attack Surfaces in Multimodal Large Language Models}
Multimodal large language models (MLLMs) combine computationally intensive multimodal perception with autoregressive generation, thereby exposing a broader attack surface.

\subsubsection{Risks of Overthinking.}
Related risks had already appeared in earlier vision systems, that visual perturbations can weaken sparsity benefits, negate dynamic quantization gains, or overload downstream vision pipelines, as seen in Uniform Inputs~\cite{muller2024impact}, QuantAttack~\cite{baras2025quantattack}, and Phantom Sponges and Poison-splat~\cite{shapira2023phantom, schoof2024beyond, lu2024poison}. Although these works do not directly target MLLMs, they foreshadow that seemingly benign vision inputs can be crafted to induce disproportionate resource consumption.

This risk becomes more explicit in MLLMs. Existing attacks manipulate the autoregressive decoding process to trigger excessively long outputs~\cite{chen2022nicgslowdown, gao2024inducing, gao2024energy}, while more recent methods improve stealth by silently generating invisible tokens that conceal substantial computational overhead~\cite{zhang2025hidden}. Other frameworks, such as VLMInferSlow~\cite{wang2025vlminferslow} and EO-VLM~\cite{seo2025eo}, further enhance the practicality of such slowdown attacks under black-box access. Together, these studies show that in MLLMs, overthinking is a concrete attack surface that can be exploited to inflate inference cost and degrade system efficiency.

\subsubsection{Risks of Unbounded Drift.}
LingoLoop~\cite{fu2025lingoloop} and RECITE~\cite{gao2025resource} reveal a multimodal form of unbounded drift in image-conditioned generation. The former traps the model in linguistically constrained visual descriptions, while the latter induces repetitive visual recall, driving visual-language interaction into a self-reinforcing decoding loop.

\begin{tcolorbox}[
  colback=lightbox,
  colframe=myorange,
  colbacktitle=myorange,
  coltitle=white,
  fonttitle=\bfseries,
  title=Key Insights.,
  rounded corners,
  arc=3mm
]

Existing studies have revealed significant vulnerabilities in MLLMs, but most focus on single-modal attacks, with limited attention to cross-modal threats. Furthermore, the potential risks in audio and video modalities have not been fully revealed.

\end{tcolorbox}

\subsection{Attack Surfaces in Agentic Systems}

Agentic systems enable large language models to interact with real-world environments, but they also introduce new attack surfaces that may increase real-world risks.
LLM-based agent systems have also been shown to be vulnerable to resource consumption attacks. LeechHijack \cite{zhang2025leechhijackcovertcomputationalresource} injects auxiliary workloads into an agent’s reasoning loop, covertly steering it to perform attacker-specified computation while maintaining seemingly normal outputs. CORBA \cite{zhou2025corbacontagiousrecursiveblocking} targets multi-agent systems by propagating self-replicating prompts across agents, inducing recursive interactions that waste computational resources and undermine system availability.

Computer-use agents such as OpenClaw~\cite{openclaw2026} and Cloud Code~\cite{anthropic2025claudecode_repo} are designed to interact directly with operating systems. During execution, these agents may spawn persistent background processes without clear termination conditions, leading to prolonged server-side resource consumption \cite{shapira2026agentschaos}.

Beyond computer-use settings, related resource consumption risks have also been observed in other domains. In autonomous driving, existing attacks primarily target perception pipelines, where adversaries increase detection proposals or tracked objects to induce inference latency \cite{chen2021transslowdown, Chen_2024_CVPR,11023507,Ma_Wang_Chen_Shen_2024}. CP-FREEZER \cite{wang2025cpfreezerlatencyattacksvehicular} focuses on cooperative perception scenarios, where perturbations applied to shared features increase the computational overhead of non-maximum suppression (NMS). SlowLiDAR \cite{Liu_2023_CVPR} further extends resource consumption from the visual modality to the point cloud modality. Agents deployed on edge devices are likewise vulnerable to attacks that significantly increase energy consumption and inference latency \cite{hasan2025spongeattackssensingai}.

\begin{tcolorbox}[
  colback=lightbox,
  colframe=myorange,
  colbacktitle=myorange,
  coltitle=white,
  fonttitle=\bfseries,
  title=Key Insights.,
  rounded corners,
  arc=3mm
]

Existing studies reveal vulnerabilities in resource consumption within agentic systems, yet analysis of resource use across agent components remains limited. In particular, the resource consumption characteristics of agent memory mechanisms are still largely unexplored. Moreover, many downstream applications of agentic systems have not been systematically studied from the perspective of resource consumption.

\end{tcolorbox}

\section{Mechanisms Underlying Resource Consumption Behaviors}
\label{sec:mechanisms}


This section reviews mechanistic and interpretability research on the emergence of resource-intensive behaviors in autoregressive generation.
Additional technical details are provided in Appendix~\ref{sec:appen_mechanisms}.

\subsection{Mechanisms Behind Overthinking}

Mechanistic analysis of overthinking remains limited. Unlike unbounded drift, overthinking often stays task-relevant and may appear as an extreme extension of otherwise normal generation, making its causal boundary harder to isolate. As a result, existing studies provide only partial clues rather than a mature taxonomy of mechanisms. Prior work suggests that unnecessarily prolonged generation may arise from both probability-level trapping effects, such as high-inflow tokens that bias decoding toward extended trajectories \cite{fu2021theoretical}, and attention irregularities that drive repeated divergence into loosely related continuations \cite{pmlr-v267-yona25a}. Together, these findings offer preliminary evidence that overthinking may be sustained by local decoding dynamics even when generation remains superficially coherent.

\begin{tcolorbox}[
  colback=lightbox,
  colframe=myorange,
  colbacktitle=myorange,
  coltitle=white,
  fonttitle=\bfseries,
  title=Key Insights.,
  rounded corners,
  arc=3mm
]

Output prolongation stems from attention divergence and probability traps. However, current interpretability research focuses on simple failures, leaving verbose reasoning and overthinking in Chain-of-Thought settings largely underexplored.

\end{tcolorbox}

\subsection{Mechanisms Behind Unbounded Drift}

\paragraph{Theoretical Foundations.}
Early studies on degenerative generation provide initial clues about the roots of unbounded drift. Prior work suggests that autoregressive decoding can be drawn into repetitive trajectories by attractor-like token dynamics or collapse toward a limited subset of tokens \cite{fu2021theoretical,ildiz2024self}. Related evidence also points to training objectives as a contributing factor: unlikelihood training \cite{welleck2019neural} was introduced to suppress repetitive outputs beyond standard likelihood optimization, while subsequent work argues that such auxiliary penalties may still be insufficient to fully correct the underlying probability skew \cite{lin2021straight}. Together, these studies suggest that unbounded drift may reflect not only inference-time failures but also deeper biases in sequence modeling and training.

Subsequent work has moved beyond output-level penalization toward intervening at the source of repetition. HAPAX~\cite{sahin2025context}, for example, excludes induction-head-predictable tokens from the training objective, while Repetition Dropout~\cite{li2023repetition} reduces reliance on repeated context by randomly dropping attention to repeated words during training. Taken together, these studies suggest that unbounded drift is shaped not only by inference-time dynamics but also by the objectives and optimization signals used during training. 
Notably, however, this line of analysis has so far been developed primarily for text-based LLMs, with a much less mechanistic understanding of reasoning or multimodal settings.

\paragraph{Circuit and Representation Mechanisms.}
From a mechanistic perspective, unbounded drift appears to arise not from a single faulty component, but from coordinated repetition-promoting circuits spanning attention heads, downstream neurons, and latent features. Prior work first identifies repetition-related neurons distributed across layers, with intermediate layers detecting repeated patterns and higher layers amplifying context copying~\cite{hiraoka2025repetition}. Layer-wise causal analysis further suggests that these neurons do not operate in isolation, but often function downstream of induction heads that propagate copying signals~\cite{doan2025understanding}. Moving beyond individual units, Sparse Autoencoder analyses reveal latent directions that explicitly encode repetition-related features~\cite{yao-etal-2025-understanding}, indicating that repetitive behavior is represented at the feature level rather than emerging as a purely surface-level artifact. This view is further reinforced by studies connecting induction heads to in-context learning and repetition dynamics~\cite{crosbie2025induction,wang2025induction,mahaut2025repetitions}. Taken together, these findings suggest that unbounded drift reflects a structured internal bias toward copying and repetition, which can be amplified into persistent looping once normal stopping behavior breaks down.

\textbf{Behavioral Dynamics.}
At the behavioral level, repetition exhibits strong self-reinforcing dynamics once triggered.
Pseudo-Repetition Penalization~\cite{xu2022learning} shows that, after repetition begins, the probability of generating identical content rapidly increases, forming a closed loop.
Complementarily, uncertainty-driven fallback modeling~\cite{ivgi2024loops} interprets repetition as a fallback behavior under uncertainty, where models revert to low-entropy states by repeating prior text when unable to identify a plausible continuation.

\begin{tcolorbox}[
  colback=lightbox,
  colframe=myorange,
  colbacktitle=myorange,
  coltitle=white,
  fonttitle=\bfseries,
  title=Key Insights.,
  rounded corners,
  arc=3mm
]

Unbounded drift stems from both model architecture and training methods. While decoding-time interventions offer immediate relief, training-level approaches may provide more stable, fundamental mitigation by addressing root causes rather than surface-level symptoms.

\end{tcolorbox}

\section{Mitigation Strategies for Resource Consumption Threats}
\label{sec:evaluation}

This section reviews existing mitigation methods for threats to resource consumption.
Importantly, these methods are not equally security-oriented: many aim to improve efficiency under benign conditions, whereas only a smaller subset explicitly defends against adversarial resource amplification.
Detailed technical descriptions of mitigation strategies are provided in Appendix~\ref{sec:appendix_defense}.

\subsection{Mitigating Overthinking}

\paragraph{LLMs and RLLMs.}
For reasoning models, most existing mitigations are better understood as efficiency-oriented interventions rather than defenses explicitly designed against adversarial resource amplification. Existing overthinking is often studied as an inference-efficiency problem; current methods mainly aim to reduce unnecessary reasoning length under benign conditions, while only indirectly mitigating attack-induced overhead.

At the training level, these methods aim to foster awareness of the length of reasoning. 
\citet{wu2025lapo} introduce budget signals through reward shaping, while parameter-space tuning identifies controllable directions that compress reasoning chains within a single model~\citep{ma-etal-2025-cot}.

At the decoding level, lightweight interventions suppress redundant reasoning without retraining, for example, by stopping unnecessary reflection once the model is sufficiently certain or by interrupting repetitive hidden-state patterns and injecting control vectors to regulate thinking depth~\citep{huang2025cgrs,xie-etal-2025-word, liu2025selfaffirmation, lin2025controlling, he2025frponder}.

At the system level, resource allocation is adjusted more explicitly. Difficulty-adaptive methods allocate inference budget according to query complexity~\citep{wang-etal-2025-make}, while token-budget-aware prompting and monitoring-of-thought frameworks terminate unnecessary reasoning externally without modifying model parameters~\citep{han-etal-2025-token,zhu2025unthinking}.

Overall, these methods show that overthinking can be mitigated from multiple levels, but the current literature remains centered on efficiency optimization rather than attack-specific defense.

\paragraph{Multimodal Large Language Models (MLLMs).}
In multimodal settings, overthinking can be amplified by both unnecessary reasoning and large external contexts. Certainty-based routing~\citep{lu2025car} dynamically invokes CoT only when the model is uncertain, thereby avoiding prolonged reasoning on simpler inputs. Memory-augmented frameworks~\citep{bhatnmret, ottem2025meve} further mitigate context bloat through structured memory modules and verification-centric filtering, significantly reducing context size without degrading retrieval quality.

\begin{tcolorbox}[
  colback=lightbox,
  colframe=myorange,
  colbacktitle=myorange,
  coltitle=white,
  fonttitle=\bfseries,
  title=Key Insights.,
  rounded corners,
  arc=3mm
]

Existing mitigations have shown effectiveness in alleviating overthinking.
However, current methods predominantly focus on efficiency-oriented mitigations rather than attack-aware defenses, and explicit mitigations for autonomous agent pipelines remain largely unexplored.
\end{tcolorbox}

\subsection{Mitigating Unbounded Drift}

\paragraph{General LLMs.}

Unbounded drift in text generation often manifests as repetitive or uncontrolled decoding. As a result, mitigation strategies in general LLMs mainly focus on detecting or suppressing repetition to prevent outputs from degenerating into runaway generation loops.

At the model level, existing methods mitigate degeneration either through training-time objective design or decoding-time control. Unlikelihood training~\citep{welleck2019neural} reduces degenerate text by penalizing repeated tokens, while RAP~\citep{huang2025rap} provides a structured way to tune repetition penalties during decoding while maintaining task performance.

At the system level, robustness is enforced through runtime detection and scheduling safeguards. For instance, RecurrentDetector~\citep{yu2025breaking} identifies repetitive activation patterns to halt infinite loops. Additionally, PD\textsuperscript{3}F~\citep{zhang2025pd3f} combines request scheduling with output EOS amplification. Queue-based architectures also help stabilize throughput under heavy workloads~\citep{shahriar2025vulnerability}.

Overall, these methods show that uncontrolled repetitive generation in text models can be effectively mitigated through interventions across training, decoding, and system layers.

\paragraph{Reasoning Models.}
Dedicated defenses targeting non-terminating reasoning in RLLMs remain scarce. Existing techniques such as WSC~\citep{xie-etal-2025-word} and self-affirmation suppression~\citep{liu2025selfaffirmation} primarily focus on removing redundant reasoning during decoding, which can incidentally alleviate mild repetitive behaviors.

\paragraph{Agentic Systems.}
Agentic systems introduce additional vulnerabilities due to continuous perception pipelines. Background-attentive adversarial training~\citep{wang2025cantslowmedown} improves robustness by incorporating perturbation-aware training objectives, effectively restoring inference speed and stability under resource abuse on edge devices.

\begin{tcolorbox}[
  colback=lightbox,
  colframe=myorange,
  colbacktitle=myorange,
  coltitle=white,
  fonttitle=\bfseries,
  title=Key Insights.,
  rounded corners,
  arc=3mm
]

System-level frameworks effectively mitigate crash-level threats and DoS-style attacks in general LLMs. However, purpose-built crash defenses for MLLMs remain largely absent.

\end{tcolorbox}

\section{Open Challenges and Future Directions}
\label{sec:position}
\subsection{From Efficiency Optimization to Security Guarantees}

A useful distinction is between efficiency-oriented mitigation and security-oriented defense. The former treats excessive generation primarily as a cost problem under benign workloads, whereas the latter treats it as an adversarial resource-amplification threat that requires robustness guarantees.
Much of the literature still frames excessive generation as a cost-latency trade-off, aiming to reduce average output length under benign conditions rather than defend against adversarial resource amplification~\cite{jiang2024longllmlingua,agrawal2024taming,del2023skipdecode}.

Looking forward, two research directions appear especially important. First, future research should move from heuristic efficiency control to security-oriented budget protection, treating resource usage as a constrained attack surface rather than merely an optimization target. Second, defenses should protect not only output-level resource usage but also process-level computation, including intermediate reasoning, tool interactions, and multi-turn execution trajectories. Together, these directions would help elevate resource control from an efficiency concern to a security objective for reliable and sustainable LLMs.

\subsection{Understanding Threat Mechanisms}

A major limitation of current research is the lack of a unified mechanistic understanding of resource amplification. 
Existing studies largely focus on specific attack instances, such as reasoning over-expansion or image-coded induction~\cite{gao2025resource, fu2025lingoloop}, while offering little general theory of how generation trajectories expand under attack manipulation. Future work should therefore integrate the shared dynamics and model-specific properties of different generative systems to develop transferable theories that support cross-model analysis and principled defense design.

\subsection{Resource Abuse in LLM Ecosystems}
LLMs expose an increasingly serious form of resource abuse, yet existing studies remain limited to a small number of attack surfaces~\cite{zhang2025leechhijackcovertcomputationalresource,zhou2026beyond}. Compared with model-level attacks, these threats are harder to characterize because resource consumption may accumulate across tool calls, intermediate services, memory updates, and multi-agent coordination, while the final output can still appear normal~\cite{lee2026overthinking}. 
Future research should therefore move beyond isolated case studies toward a systematic understanding of agent-level resource abuse, including component-wise resource accounting, cross-step dependency analysis, and defenses that can detect or interrupt malicious amplification throughout execution. 
This progress is increasingly important as agentic systems are deployed in real-world workflows, where resource abuse can directly undermine the reliability and sustainability of the services they support.

\subsection{Toward Standardized Evaluation}

A fundamental limitation of current research is the lack of a standardized evaluation framework for resource consumption threats. Existing studies span different model families, modalities, and deployment settings, but they often rely on heterogeneous metrics, which makes cross-study comparison difficult. 
To address this gap, Appendix~\ref{sec:appen_evaluation} summarizes existing evaluation practices and presents a more unified evaluation framework.
In particular, we argue that resource consumption threats should be assessed jointly from model-level behavior, hardware-level pressure, and application-level service impact, so that attacks and defenses can be compared under a common view. 
Without such a shared protocol, it remains difficult to consistently measure amplification severity, comprehensively evaluate defense coverage, or support downstream governance of resource consumption risks.

\section{Conclusion}

Resource consumption threats are emerging as an important safety concern for large language models. By inducing excessive and unnecessary generation, they not only reduce efficiency but also create broader risks for system reliability, service availability, and operational cost. This survey provided a unified view of the area through two representative regimes, \textit{Overthinking} and \textit{Unbounded Drift}, and reviewed existing work spanning attacks, mechanisms, defenses, and open challenges.
Despite recent progress, the field remains fragmented in taxonomy, theory, evaluation, and mitigation. 
We hope this survey provides a clearer foundation for future research on understanding and mitigating threats to resource consumption and helps advance safer, more sustainable LLMs.

\section*{Limitations}

This survey has several limitations. First, our discussion centers on \emph{resource consumption threats in LLMs}, rather than the broader landscape of efficiency, robustness, or availability problems. In particular, we primarily focus on malicious or adversarial resource abuse that induces excessive generation, and therefore do not aim to comprehensively cover benign efficiency optimization, general systems engineering for acceleration, or all forms of denial-of-service behavior outside the generative process itself. As a result, some closely related work on inference acceleration, scheduling, or hardware optimization is discussed only when it directly informs the security perspective of resource consumption.

Second, the taxonomy proposed in this survey is intended as a unifying conceptual abstraction rather than a complete partition of all possible failure modes. While this distinction helps organize existing studies by the evolution pattern of generation, some attacks may exhibit mixed characteristics, shift between the two regimes across settings, or involve multiple system components simultaneously. This issue is especially relevant in agentic systems, where resource amplification may accumulate across tool calls, memory updates, retrieval steps, and multi-agent coordination, making strict categorization more difficult.

Third, the literature in this area remains highly uneven across model families and attack settings. Existing studies are concentrated on text-based LLMs and a relatively small set of open-source models, while black-box commercial systems remain harder to study systematically. Coverage is also imbalanced across modalities and deployment scenarios: current multimodal work still focuses heavily on image-centered settings, whereas the risks in audio and video modalities remain far less explored; similarly, many downstream agent applications have not yet been systematically examined from the perspective of resource consumption.

Finally, because this is an emerging field, the available empirical evidence remains fragmented, and evaluation practices are not yet standardized. Many papers report attack success using different metrics, threat models, and deployment assumptions, making direct comparison difficult and potentially limiting the stability of any unified conclusions. For this reason, our synthesis should be understood as a structured overview of the current research landscape rather than a definitive benchmark of attack prevalence, mechanism universality, or defense effectiveness.

\section*{Ethical Considerations}

This survey reviews resource consumption attacks on LLMs. Although summarizing attack mechanisms may increase understanding of such threats, our goal is strictly defensive: to clarify the threat landscape, support standardized evaluation, and encourage effective mitigation. We therefore focus on conceptual analysis and system-level implications rather than actionable attack instructions.

We further emphasize that resource consumption abuse can have real-world consequences beyond efficiency degradation. In shared LLM infrastructure, excessive generation may reduce service availability, increase operational costs, and harm end users through higher latency, unstable access, and higher usage costs. These risks can become even more severe in agentic systems, where recursive execution and covert workload amplification may disrupt workflows and cause hidden financial loss. By synthesizing this emerging area, we aim to support safer, more reliable, and more sustainable LLMs.


\bibliography{custom}

\appendix

\section{Scope and Inclusion Criteria}

This survey focuses on threats to resource consumption in LLMs. Our primary focus is on adversarial or malicious resource abuse that increases computational cost through extended, excessive, or uncontrolled generation. Accordingly, we treat attacks on LLMs, reasoning models, multimodal generative systems, and agentic pipelines as the core literature when they induce abnormal output length, non-convergent decoding, recursive execution expansion, or related forms of resource amplification.

We also discuss a limited set of closely related works from vision and systems security, including selected latency and sponge-style attacks, but only in a supporting role. These works are included when they provide historical precursors, analogous threat patterns, or mechanistic insights that help explain later resource abuse in generative systems. They should not be interpreted as equally central instances of the problem studied in this survey.

By contrast, we do not aim to comprehensively review benign efficiency optimization, generic inference acceleration, hardware scheduling, or all forms of denial-of-service and availability attacks outside the generative process itself. Such work is referenced only when it directly informs the security perspective on resource consumption threats or helps contextualize the survey's boundary.

\section{Preliminary}
\label{sec:appen_preliminary}
\subsection{Autoregressive Generation}

Most modern LLMs produce outputs through an autoregressive decoding process~\cite{vaswani2017attention,malach2024auto}.
Given an input prompt $x$, the model generates an output sequence $y = (y_1, y_2, \dots, y_T)$ token by token according to the conditional distribution:
\begin{equation}
p(y_t \mid x, y_{<t}),
\end{equation}
where $y_{<t}$ denotes the previously generated tokens and $T$ is the final generation length.
At each decoding step, the model computes a probability distribution over the vocabulary conditioned on the prompt and previously generated tokens, and selects the next token via sampling or deterministic decoding.

The decoding process continues iteratively until a termination condition is satisfied.
Common termination criteria include emitting a special end-of-sequence token, reaching a predefined maximum length, or satisfying task-specific stopping rules.
Consequently, the overall generation process can be viewed as a sequence of decoding steps, each extending the partial output and updating the model's internal state.

Because each token requires an additional decoding step, the generation length $T$ directly determines the number of forward passes executed during inference.
As a result, the computational cost of a generation process grows with the length of the generated sequence.
This characteristic makes generation length a central factor in determining the resource usage of modern LLMs.

\subsection{Computational Cost of Generation}

The computational cost of autoregressive generation arises from repeated forward passes through large neural networks.
For a model with parameters $\theta$, generating a sequence of length $T$ requires evaluating the model $T$ times to compute the conditional probabilities $p(y_t \mid x, y_{<t}; \theta)$.

In addition to repeated model evaluation, attention-based architectures introduce further computational overhead.
Self-attention mechanisms require each token to attend to all previously generated tokens, leading to increased memory usage and computation as the sequence length increases.
Consequently, the total computational cost of generation can be approximated as a function of the output length:
\begin{equation}
C(T) = \sum_{t=1}^{T} c_t,
\end{equation}
where $c_t$ denotes the cost of the $t$-th decoding step.
In practice, $c_t$ may grow as the context window expands, leading to increasing latency and memory consumption for longer sequences.

These computational characteristics directly translate into operational costs in real-world deployments.
Longer generation sequences lead to higher GPU utilization, increased inference latency, and greater energy consumption.
For this reason, many commercial LLM services use token-based pricing models, charging users based on the number of generated tokens.
Under such pricing schemes, longer outputs correspond directly to higher economic costs.

Therefore, generation length becomes a critical resource variable in LLMs.
Any behavior that unnecessarily increases the number of generated tokens may significantly amplify computational cost and degrade system throughput.

\subsection{Generation Trajectories}

The decoding process can be viewed as a trajectory that describes how the model progresses from an input prompt to a final output.
Formally, a generation process can be represented as a trajectory:
\begin{equation}
\tau = (x, y_1, y_2, \dots, y_T),
\end{equation}
which evolves over successive decoding steps.

Under normal conditions, the trajectory converges once the model produces a complete response that satisfies the task objective.
However, the dynamics of this trajectory may be altered by adversarial prompts, system-level manipulations, or unexpected interaction patterns.
In such cases, the generation trajectory may extend significantly beyond the length required for task completion.

Because computational cost is tightly coupled with the trajectory length, deviations in trajectory dynamics can directly amplify resource consumption.
Understanding how generation trajectories evolve is therefore essential for analyzing resource consumption behaviors in LLMs.

\section{The Rapid Growth of Context and Cost in LLM}
\begin{table}[t]
\centering
\resizebox{\columnwidth}{!}{%
\begin{tabular}{@{}l|c@{}}
\toprule
\textbf{Model}\quad\quad & \quad\quad\textbf{Context Window}\quad\quad \\ \midrule
\rowcolor[HTML]{EFEFEF} 
GPT-5.4 / GPT-5.4-pro & 1M \\ \midrule
Claude Opus 4.6 & 200K (1M beta) \\
\rowcolor[HTML]{EFEFEF} 
Claude Sonnet 4.6 & 200K (1M beta) \\ \midrule
Gemini 2.5 Pro & 1M \\
\rowcolor[HTML]{EFEFEF} 
Gemini 2.5 Flash & 1M \\ \midrule
Mistral Large 2 & 128K \\ \midrule
\rowcolor[HTML]{EFEFEF} 
DeepSeek-V3 / R1 & 128K \\ \bottomrule
\end{tabular}%
}
\caption{Historical Growth of Context Windows in Mainstream and Frontier Models}
\label{tab:model_len}
\end{table}
\begin{table*}[t]
\centering
\begin{threeparttable}
\resizebox{0.7\textwidth}{!}{%
\begin{tabular}{@{}l|cc@{}}
\toprule
\cellcolor[HTML]{FFFFFF}{\color[HTML]{282828}\textbf{Model}}\quad\quad &
\quad\quad\textbf{Input}\quad\quad &
\quad\quad\textbf{Output}\quad\quad \\
\midrule
\rowcolor[HTML]{EFEFEF}
{\color[HTML]{282828} GPT-5.4 (\textless{}272K context length)\tnote{a}} & {\color[HTML]{282828} \$2.50} & {\color[HTML]{282828} \$15.00} \\
{\color[HTML]{282828} GPT-5.4 ($>$272K context length)\tnote{a}} & {\color[HTML]{282828} \$5.00} & {\color[HTML]{282828} \$22.50} \\
\rowcolor[HTML]{EFEFEF}
{\color[HTML]{282828} GPT-5.4-pro (\textless{}272K context length)\tnote{a}} & {\color[HTML]{282828} \$30.00} & {\color[HTML]{282828} \$180.00} \\
{\color[HTML]{282828} GPT-5.4-pro ($>$272K context length)\tnote{a}} & {\color[HTML]{282828} \$60.00} & {\color[HTML]{282828} \$270.00} \\
\rowcolor[HTML]{EFEFEF}
{\color[HTML]{282828} GPT-5 mini\tnote{a}} & \$0.25 & \$2.00 \\
\midrule
{\color[HTML]{282828} Claude Opus 4.6\tnote{b}} & {\color[HTML]{3D3D3A} \$5.00} & {\color[HTML]{3D3D3A} \$25.00} \\
\rowcolor[HTML]{EFEFEF}
{\color[HTML]{282828} Claude Sonnet 4.6\tnote{b}} & {\color[HTML]{3D3D3A} \$3.00} & {\color[HTML]{3D3D3A} \$15.00} \\
{\color[HTML]{282828} Claude Haiku 4.5\tnote{b}} & {\color[HTML]{3D3D3A} \$1.00} & {\color[HTML]{3D3D3A} \$5.00} \\
\midrule
\rowcolor[HTML]{EFEFEF}
{\color[HTML]{282828} Gemini 2.5 Pro ($\leq$200K prompt)\tnote{c}} & \$1.25 & \$10.00 \\
{\color[HTML]{282828} Gemini 2.5 Pro ($>$200K prompt)\tnote{c}} & \$2.50 & \$15.00 \\
\rowcolor[HTML]{EFEFEF}
{\color[HTML]{282828} Gemini 2.5 Flash\tnote{c}} & \$0.15 & \$1.25 \\
{\color[HTML]{282828} Gemini 2.5 Flash-Lite\tnote{c}} & \$0.10 & \$0.40 \\
\rowcolor[HTML]{EFEFEF}
{\color[HTML]{282828} Gemini 2.0 Flash\tnote{c}} & \$0.05 & \$0.20 \\
\midrule
{\color[HTML]{282828} Mistral Large 2\tnote{d}} & \$2.00 & \$6.00 \\
\rowcolor[HTML]{EFEFEF}
{\color[HTML]{282828} Mistral Medium 3 / 3.1\tnote{d}} & \$0.40 & \$2.00 \\
{\color[HTML]{282828} Mistral Small\tnote{d}} & \$0.20 & \$0.60 \\
\midrule
\rowcolor[HTML]{EFEFEF}
{\color[HTML]{282828} DeepSeek-V3\tnote{e}} & $\sim$\$0.27 & $\sim$\$1.1 \\
{\color[HTML]{282828} DeepSeek-R1\tnote{e}} & $\sim$\$0.55 & $\sim$\$2.2 \\
\bottomrule
\end{tabular}%
}
\begin{tablenotes}
\footnotesize
\item[a] OpenAI pricing page: \url{https://developers.openai.com/api/docs/pricing}
\item[b] Anthropic pricing page: \url{https://platform.claude.com/docs/en/about-claude/pricing}
\item[c] Google AI pricing page: \url{https://ai.google.dev/pricing}
\item[d] Mistral pricing page: \url{https://mistral.ai/pricing\#api}
\item[e] DeepSeek pricing page: \url{https://api-docs.deepseek.com/quick_start/pricing/}
\end{tablenotes}
\caption{API pricing comparison of representative mainstream models. Prices are listed in USD per 1M input/output tokens.}
\label{tab:api_cost}
\end{threeparttable}
\end{table*}

Recent LLMs have expanded rapidly in both capability and serving scale, with context windows and deployment budgets increasing in parallel. As shown in Table~\ref{tab:api_cost}, current mainstream APIs already exhibit substantial variation in usage costs, especially for high-end, reasoning-oriented, and frontier models, whose output-token pricing remains significantly higher than that of lightweight alternatives. At the same time, Table~\ref{tab:model_len} shows that the context windows of representative commercial models have reached 128K, 200K, and even 1M tokens, making long-context generation an increasingly common operating regime rather than a special-case setting. This trend is further reflected in Table~\ref{tab:year_len}, where both mainstream and frontier models exhibit rapid year-over-year growth in supported context length.

Taken together, these trends indicate that modern LLMs are no longer optimized solely for short-response interactions but are increasingly designed to support long-horizon reasoning, retrieval, and multi-step execution. As context windows and output budgets continue to expand, the economic upper bound of a single interaction also rises accordingly. Under such conditions, any unnecessary expansion in reasoning length, retrieval scope, or execution trajectory can directly translate into higher API expenditure and a larger system-side computational burden. This makes resource consumption analysis increasingly important, not only for understanding attack surfaces but also for characterizing the practical cost environment in which such threats emerge.
\begin{table*}[t]
\centering
\resizebox{0.8\textwidth}{!}{%
\begin{tabular}{@{}l|cccccc@{}}
\toprule
\textbf{Models}\quad\quad & \quad \textbf{2020}\quad\quad & \quad \textbf{2021}\quad\quad & \quad \textbf{2022}\quad\quad & \quad \textbf{2023}\quad\quad & \quad \textbf{2024}\quad\quad & \quad \textbf{2025}\quad\quad \\ \midrule
\rowcolor[HTML]{EFEFEF} 
Mainstream Models & 0.5K & 2K & 8K & 32K & 128K & 128K \\
Frontier Models & 1K & 4K & 16K & 128K & 512K & 10M \\ \bottomrule
\end{tabular}%
}
\caption{Context Window Comparison of Representative Mainstream Models}
\label{tab:year_len}
\end{table*}

\section{The Resource Consumption Threat Landscape}
\label{sec:appen_threat}

\subsection{The Growing Pressure on Computational Resources}

In LLM systems, computational cost is tightly coupled with the generation process itself: longer outputs require more autoregressive decoding steps~\cite{vaswani2017attention} and impose growing sequence-length-dependent memory and compute overhead~\cite{nayab2025concise,wang2024minions,anagnostidis2023dynamic}. 
As a result, resource pressure becomes an inherent property of modern generation systems.
This relationship is already reflected in real-world deployment practice~\cite{lodha2025tokenops}, where providers charge separately for output tokens and offer discounted asynchronous batch services to better manage compute demand\footnote{\url{https://openai.com/zh-Hans-CN/api/pricing/?utm_source=chatgpt.com}}. In large-scale deployments, such pressure can escalate into visible service risks; for example, the launch of DeepSeek-R1 attracted massive traffic, while the cost of long-form generation contributed to severe capacity strain and service restrictions\footnote{\url{https://www.reuters.com/technology/cybersecurity/deepseek-limits-registrations-due-cyber-attack-2025-01-27/?utm_source=chatgpt.com}}. These observations suggest that the generation process itself has become a bottleneck in modern model deployment, making resource-amplifying behaviors a critical concern for reliable and sustainable deployment.

\subsection{Threat Scenarios Under Resource Consumption Regimes}

Building on the taxonomy introduced in the main text, resource consumption threats can manifest under two representative regimes: \textbf{\textit{Overthinking}} and \textbf{\textit{Unbounded Drift}}. Although both ultimately aim to amplify computational consumption during generation, they differ in the behavioral patterns of the generation process and in the types of risks they introduce.

\textbf{Overthinking threats.}
In the Overthinking regime, the generation process remains semantically aligned with the task but becomes unnecessarily verbose or computationally intensive. Such behavior can arise when prompts or inputs induce the model to produce excessively detailed reasoning~\cite{zhu2025extendattackattackingserverslrms}, redundant explanations~\cite{kumar2025overthinkslowdownattacksreasoning}, or low-information-density descriptions~\cite{shumailov2021sponge}. This phenomenon is particularly prominent in reasoning-oriented models~\cite{wei2022chain,creswell2022faithful,hao2023reasoning,guo2025deepseek,he2025skywork} and agent-based systems~\cite{yao2022react,xi2025rise,wang2024survey,chen2023autoagents}, where extended reasoning chains or repeated tool interactions may significantly expand the generation trajectory. For example, adversarial manipulations of reasoning prompts can induce models to generate prolonged chains of intermediate reasoning~\cite{si2025excessivereasoningattackreasoning}, dramatically increasing token consumption and latency~\cite{yi2025badreasonerplantingtunableoverthinking}. Similar amplification effects have also been observed in multimodal systems, where carefully crafted inputs can trigger extremely verbose textual outputs despite containing limited semantic information~\cite{gao2025resource}.
These behaviors may appear benign at the output level, yet they substantially inflate inference cost and degrade overall system throughput. When scaled across large numbers of queries, such amplification can accumulate into significant economic losses or operational overhead. Several realistic deployment scenarios illustrate how such behavior can be exploited in practice. 

First, in commercial LLM services, a malicious or poorly regulated service provider could intentionally design models or prompting pipelines that encourage unnecessarily verbose reasoning or redundant explanations, effectively increasing token consumption and inflating user-side API costs~\cite{zhu2025unthinking}. Because users typically pay for output tokens~\cite{alavi2025cost}, even subtle increases in verbosity can translate into substantial financial overhead when deployed at scale~\cite{sun2025coin,lodha2025tokenops,bergemann2025economics}.

Second, in agent-based ecosystems where multiple tools, plugins, or intermediate services participate in a task pipeline, adversarial intermediaries may inject additional reasoning steps, auxiliary tasks, or low-value instructions into the agent workflow~\cite{zhan2024injecagent,zhao2025mind,zhang2025leechhijack}. Such manipulations can cause the model to perform extra computations that appear semantically related to the task while silently consuming additional API budget or computational resources~\cite{dong2026clawdrainexploitingtoolcallingchains,wang2025draincode}.

Third, overthinking behaviors can pose availability risks for resource-constrained deployments, such as small service providers or edge devices~\cite{Zheng3719664,reddi2025generative,fung2025embodied}. In these settings, excessive generation may significantly delay task completion, block concurrent requests, or exhaust limited compute capacity, thereby reducing system responsiveness and degrading overall service reliability~\cite{wang2023energy,chen2021transslowdown}.

Taken together, these scenarios suggest that seemingly benign verbosity in generation can become a practical vector for resource abuse when deployed in real-world LLM ecosystems.

\textbf{Unbounded Drift}, in contrast, refers to generation trajectories that progressively deviate from the intended task and fail to converge within a reasonable progression. Such behavior can arise when adversarial prompts~\cite{gao2025resource}, decoding perturbations~\cite{geiping2024coercing}, or recursive reasoning~\cite{wang2026recur}. This phenomenon is particularly prominent in autoregressive decoding and agentic systems~\cite{wang2023voyager}, where weakened termination signals, repetitive token dynamics, or recursive tool use may cause the generation trajectory to expand far beyond normal task completion. For example, carefully crafted malicious inputs can trap models in repetitive decoding loops or non-halting generation, forcing them to continue producing tokens until system-imposed limits intervene~\cite{gao2025resource}. Similar effects can also emerge in agent environments, where recursive tool-calling or multi-step interaction chains expand the execution trajectory far beyond the original task scope~\cite{zhou2026beyond,lee2026overthinking}.
These behaviors are more overtly destructive than Overthinking, since they not only inflate computational cost but also directly threaten service availability and task completion. Several realistic deployment scenarios illustrate how such behavior can be exploited in practice.

First, malicious users or competing service providers may deliberately craft requests that suppress termination or trigger repetitive generation, forcing a provider to allocate disproportionate compute to a small number of adversarial queries~\cite{gao2024denial}. When amplified at scale, such requests can monopolize shared resources, degrade service availability, and in extreme cases resemble denial-of-service attacks against generative LLM infrastructure~\cite{li2025thinktrap,wang2026rethinking}. Recent work explicitly frames this threat as inference-time DoS, where a small malicious input can monopolize GPU time, queue slots, or memory resources, starving legitimate users~\cite{zhang2025crabs}.

Second, untrusted intermediaries in user-facing pipelines may inject malicious suffixes~\cite{zhao2025mind,si2023mondrian}, hidden instructions, or adversarial constructions into otherwise benign requests, causing the model to enter prolonged decoding trajectories or non-convergent loops. In this setting, the immediate consequence is reduced end-user usability, including longer wait times, unstable responsiveness, and higher usage costs. This threat is especially concerning because the injected content may remain invisible to end users while still shifting the request into a pathologically expensive generation regime.

Third, in agentic environments, adversarial intermediaries or compromised components may induce recursive tool-calling loops or self-amplifying interaction chains that both block task execution and corrupt task usefulness. Such behaviors not only consume excessive resources but may also trap the agent in stalled workflows, produce low-value or erroneous outputs, and obstruct downstream processes that depend on timely completion~\cite{lee2026overthinking,zhou2025corbacontagiousrecursiveblocking}. Recent studies show that tool-layer manipulation can expand agent trajectories to more than 60{,}000 tokens~\cite{zhou2026beyond}, inflate costs by hundreds of times, and substantially reduce co-running throughput, while other attacks can covertly parasitize the user’s compute budget by injecting unauthorized auxiliary workloads into apparently legitimate workflows~\cite{zhang2025leechhijackcovertcomputationalresource}. 

Taken together, these scenarios suggest that Unbounded Drift is not merely a generation abnormality, but a practical route through which resource abuse can escalate into service disruption, task failure, and system-level availability degradation.

\subsection{Why Resource Consumption Security Matters}

The preceding analysis highlights that resource consumption is no longer merely an efficiency issue but an emerging security concern in LLMs. Because modern model services operate on shared computational infrastructure, excessive generation or non-convergent interaction patterns can disproportionately consume limited resources and degrade service availability for other users. As LLMs continue to scale and become embedded in agentic workflows and real-world deployments, such behaviors may amplify operational costs, disrupt service reliability, and threaten the sustainability of model infrastructure. Despite these risks, existing studies remain fragmented across different models and system settings, lacking unified taxonomies. A clearer conceptual framework for understanding threats to resource consumption is therefore essential for advancing reliable and sustainable LLMs.

\section{A Unified View of the Survey Organization}
\label{sec:appen_survey}
\tikzstyle{my-box}= [
    rectangle,
    draw=hidden-draw,
    rounded corners,
    text opacity=1,
    minimum height=1.5em,
    minimum width=5em,
    inner sep=2pt,
    align=center,
    fill opacity=.5,
]
\tikzstyle{leaf}=[my-box, minimum height=1.5em,
    fill=blue!15, text=black, align=left,font=\large,
    inner xsep=2pt,
    inner ysep=4pt,
]
\definecolor{hidden-draw}{RGB}{20,20,20} 
\begin{figure*}[th]
    \centering
    \resizebox{\textwidth}{!}{
        \begin{forest}
            forked edges,
            for tree={
                grow=east,
                reversed=true,
                anchor=base west,
                parent anchor=east,
                child anchor=west,
                base=left,
                font=\Large,
                rectangle,
                draw=hidden-draw,
                rounded corners,
                align=left,
                minimum width=4em,
                edge+={darkgray, line width=1pt},
                s sep=3pt,
                inner xsep=2pt,
                inner ysep=3pt,
                ver/.style={rotate=90, child anchor=north, parent anchor=south, anchor=center},
            },
            where level=1{text width=7.9em,font=\Large,, align=left}{},
            where level=2{text width=8.9em,font=\large,}{},
            where level=3{text width=6.4em,font=\large,}{},
            where level=4{text width=6.4em,font=\large,}{},
        [ \;\;Resource Consumption Threat\;\; , ver
                [\;\;Attacks\\\;\;(\S \ref{sec:attack})
                    [Overthinking,text width=6em
                        [Large Language Model,text width=17.5em
                            [Sponge Poisoning~\cite{cina2025energy}; Crabs~\cite{zhang2025crabs};\\ ThinkTrap~\cite{li2025thinktrap}; BitHydra~\cite{yan2025bithydra};\\ RepetitionCurse~\cite{huang2025repetitioncursemeasuringunderstandingrouter}.
                            ,leaf ,text width=40em]
                        ]
                        [Multimodal Large Language Model ,text width=17.5em
                            [NICGSlowDown~\cite{chen2022nicgslowdown} ;Verbose Image~\cite{gao2024inducing};\\Verbose Samples~\cite{gao2024energy};Uniform Inputs~\cite{muller2024impact};\\QuantAttack~\cite{baras2025quantattack};Phantom Sponges~\cite{shapira2023phantom};\\Poison-splat~\cite{lu2024poison};Hidden Tail~\cite{zhang2025hidden};\\VLMInferSlow~\cite{wang2025vlminferslow};EO-VLM~\cite{seo2025eo};\\Enhanced PhantomSponges~\cite{schoof2024beyond}
                            ,leaf ,text width=40em]
                        ]
                        [Large Reasoning Language Model ,text width=17.5em
                            [Overthinking~\cite{kumar2025overthinkslowdownattacksreasoning};\\BadReasoner~\cite{yi2025badreasonerplantingtunableoverthinking};
                            Excessive Reasoning Attack~\cite{si2025excessivereasoningattackreasoning}\\ExtendAttack~\cite{zhu2025extendattackattackingserverslrms};BadThink~\cite{liu2025badthinktriggeredoverthinkingattacks}.
                            ,leaf ,text width=40em]
                        ]
                        [Agentic System ,text width=17.5em
                            [CLAWDrain~\cite{dong2026clawdrainexploitingtoolcallingchains}.
                            ,leaf ,text width=40em]
                        ]
                    ]
                    [\;Unbounded \\ \;\;\;\;\;\;Drift,text width=6em
                        [Large Language Model,text width=17.5em
                            [Non-halting Queries~\cite{hammouri2025non}; LoopLLM~\cite{li2025loopllm};\\ 
                            Coercing llms to do ~\cite{geiping2024coercing}; Engorgio~\cite{dong2024engorgio};\\ LLMEffiChecker~\cite{feng2024llmeffichecker}.
                            ,leaf ,text width=40em]
                        ]
                        [Multimodal Large Language Model ,text width=17.5em
                            [LingoLoop~\cite{fu2025lingoloop};RECITE~\cite{gao2025resource}
                            ,leaf ,text width=40em]
                        ]
                        [Large Reasoning Language Model ,text width=17.5em
                            [RECUR~\cite{wang2026recur}.
                            ,leaf ,text width=40em]
                        ]
                        [Agentic System ,text width=17.5em
                            [Overload~\cite{Chen_2024_CVPR}; CPFreezer~\cite{wang2025cpfreezerlatencyattacksvehicular};\\ SlowTrack~\cite{Ma_Wang_Chen_Shen_2024}; SlowLiDAR~\cite{Liu_2023_CVPR};\\ CORBA ~\cite{zhou2025corbacontagiousrecursiveblocking}; LeechHijack~\cite{zhang2025leechhijackcovertcomputationalresource}.
                            ,leaf ,text width=40em]
                        ]
                    ]
                ]
                [\;\;Performance \\ \;\;(\S \ref{sec:mechanisms})
                    [Overthinking,text width=6em
                        [Large Language Model,text width=17.5em
                            [High Inflow Words~\cite{fu2021theoretical};\\
                            Attention Sink Circuit Patching~\cite{pmlr-v267-yona25a}.
                            ,leaf ,text width=40em]
                        ]
                        [Multimodal Large Language Model ,text width=17.5em
                        ]
                        [Large Reasoning Language Model ,text width=17.5em
                        ]
                        [Agentic System ,text width=17.5em
                        ]
                    ]
                    [\;Unbounded \\ \;\;\;\;\;\;Drift,text width=6em
                        [Large Language Model,text width=17.5em
                            [Winner Takes All~\cite{ildiz2024self};
                             High Inflow Words~\cite{fu2021theoretical};\\
                             Repetition Neurons~\cite{hiraoka2025repetition};\\
                             Three-segment Neuron Ablation~\cite{doan2025understanding};\\
                             Induction Heads For ICL~\cite{crosbie2025induction};\\
                             Repetition Features~\cite{yao-etal-2025-understanding};
                             Induction Head Toxicity~\cite{wang2025induction};\\
                             Repetitions Are Not All Alike~\cite{mahaut2025repetitions};\\
                             Pseudo-Repetition Penalization~\cite{xu2022learning};\\
                             Uncertainty-Driven Fallback Modeling~\cite{ivgi2024loops};\\
                             Unlikelihood Training~\cite{welleck2019neural};
                             ScaleGrad~\cite{lin2021straight};\\
                             HAPAX~\cite{sahin2025context};
                             Repetition Dropout~\cite{li2023repetition}.
                            ,leaf ,text width=40em]
                        ]
                        [Multimodal Large Language Model ,text width=17.5em
                        ]
                        [Large Reasoning Language Model ,text width=17.5em
                        ]
                        [Agentic System ,text width=17.5em
                        ]
                    ]
                ]
                [\;\;Defense \\\;\;(\S \ref{sec:evaluation})
                    [Overthinking,text width=6em
                        [Large Language Model,text width=17.5em
                            [CCoT~\cite{nayab2025concise}.
                            ,leaf ,text width=40em]
                        ]
                        [Multimodal Large Language Model ,text width=17.5em
                            [CAR~\cite{lu2025car};NMRet~\cite{bhatnmret};MeVe~\cite{ottem2025meve}.
                            ,leaf ,text width=40em]
                        ]
                        [Large Reasoning Language Model ,text width=17.5em
                            [LAPO~\cite{wu2025lapo};CoT-Valve~\cite{ma-etal-2025-cot};CGRS~\cite{huang2025cgrs};\\WSC~\cite{xie-etal-2025-word};Self-Affirmation~\cite{liu2025selfaffirmation};\\Controlling Thinking Speed~\cite{lin2025controlling};FR-Ponder~\cite{he2025frponder};\\TALE~\cite{han-etal-2025-token};MoT~\cite{zhu2025unthinking};DSC~\cite{wang-etal-2025-make}.
                            ,leaf ,text width=40em]
                        ]
                        [Agentic System ,text width=17.5em
                        ]
                    ]
                    [\;Unbounded \\ \;\;\;\;\;\;Drift,text width=6em
                        [Large Language Model,text width=17.5em
                            [Unlikelihood Training~\cite{welleck2019neural};RAP~\cite{huang2025rap};\\RecurrentDetector~\cite{yu2025breaking};PD\textsuperscript{3}F~\cite{zhang2025pd3f};\\Queue-based~\cite{shahriar2025vulnerability}.
                            ,leaf ,text width=40em]
                        ]
                        [Multimodal Large Language Model ,text width=17.5em
                        ]
                        [Large Reasoning Language Model ,text width=17.5em
                            [WSC~\cite{xie-etal-2025-word}; Self-Affirmation~\cite{liu2025selfaffirmation}.
                            ,leaf ,text width=40em]
                        ]
                        [Agentic System ,text width=17.5em
                            [Can't Slow Me Down~\cite{wang2025cantslowmedown}.
                            ,leaf ,text width=40em]
                        ]
                    ]
                ]
        ]
        \end{forest}
    }
    \caption{Overall organization of resource consumption issues across attack, mechanism, and defense perspectives.}
    \label{fig:taxonomy}
\end{figure*}
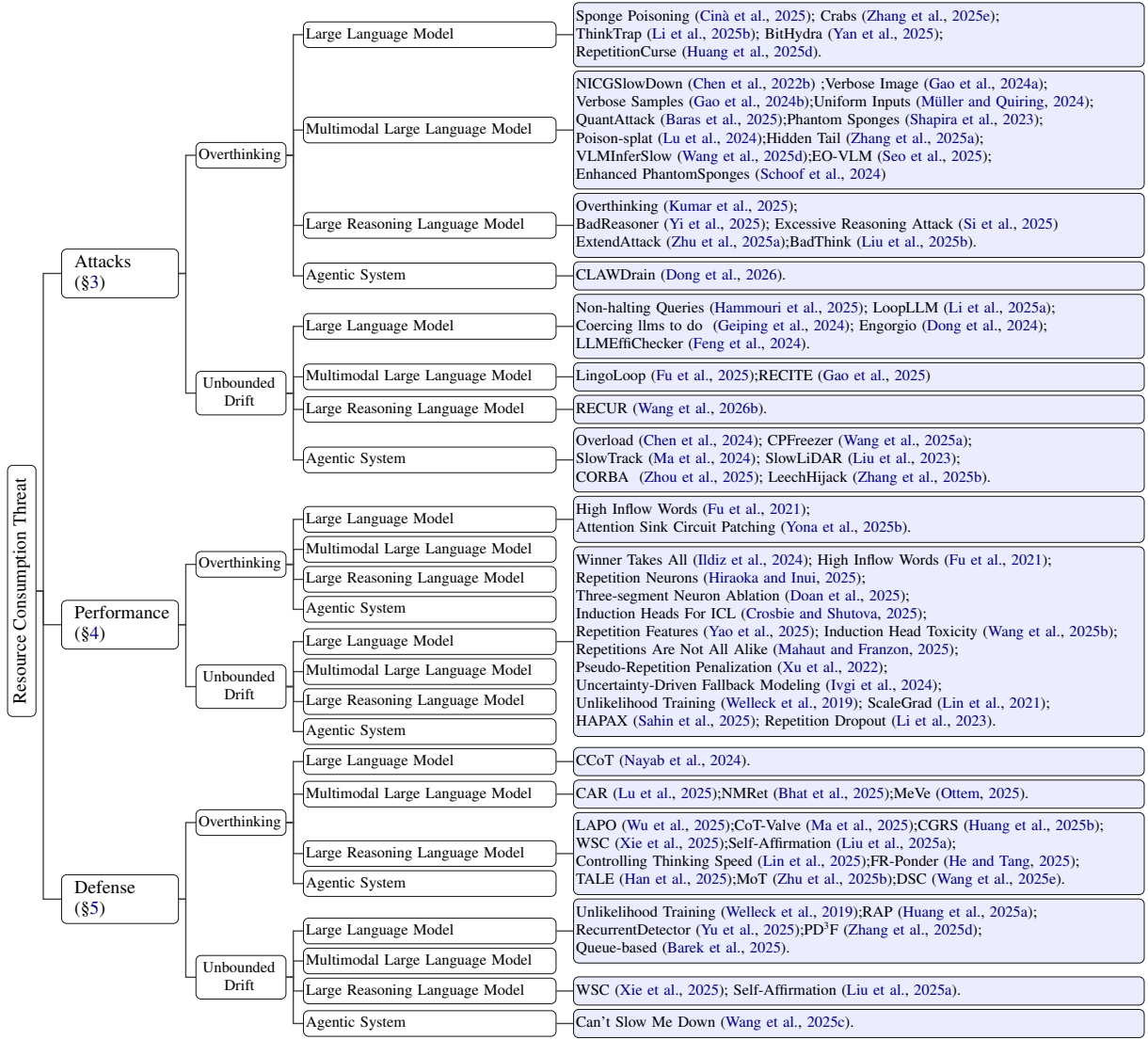
To clarify the scope and internal structure of this survey, Figure~\ref{fig:taxonomy} provides a unified overview of how the literature on resource consumption issues is organized in this work. Specifically, we structure the problem into two primary resource consumption regimes, \textit{Overthinking} and \textit{Unbounded Drift}, which serve as the main taxonomy throughout the paper. Building on this taxonomy, the survey is further developed from three complementary perspectives: \textbf{attacks}, which examine how resource amplification is induced in different systems; \textbf{mechanisms}, which analyze the underlying dynamics that lead to excessive or non-convergent generation; and \textbf{defenses}, which summarize current mitigation strategies and their coverage.

Within each perspective, existing studies are further grouped by system type, including large language models, reasoning models, multimodal large language models, and agentic systems. This organization is intended to provide a consistent view of the field: the taxonomy captures \emph{what kinds} of resource consumption behaviors emerge, while the three research perspectives capture \emph{how they are induced}, \emph{why they arise}, and \emph{how they may be mitigated}. In this way, the figure serves as a compact map of the survey, highlighting both the coverage of current research and the uneven maturity of different branches.

\section{Towards a Standardized Evaluation Framework for Resource Consumption Threats}
\label{sec:appen_evaluation}

Resource consumption threats are increasingly recognized as a critical safety concern in LLMs. While many studies have explored attacks and defenses across different modalities, model families, and deployment scenarios, the literature remains fragmented: evaluation metrics, datasets, and experimental setups vary widely, making cross-study comparison challenging. This appendix provides a structured overview of existing work and, based on observed gaps, proposes a unified evaluation framework to guide future research. The following sections first summarize the current landscape of attacks and defenses, then introduce recommended evaluation guidelines for standardized assessment.

\subsection{Summary of Existing Work}

Table~\ref{tab:attack_summary_early} to Table~\ref{tab:attack_summary_agent} present representative resource consumption attacks, including the datasets used, evaluation metrics, targeted models, and observed efficiency or latency amplification. Table~\ref{tab:deffence_summary1} and Table~\ref{tab:deffence_summary2} summarize corresponding defenses and their reported effectiveness. These tables illustrate several key points:

\textbf{High research activity across multiple modalities and settings:} Existing studies cover text-based LLMs, reasoning models, multimodal systems, and agentic environments. Attacks have been demonstrated on a wide range of models, including GPT-family, Gemini, DeepSeek, and various perception-model pipelines.

\textbf{Heterogeneous evaluation protocols:} Different works report distinct metrics, such as token amplification, latency increase, attack success rate, or system-level impact. Similarly, datasets vary from standard benchmarks to real-world agent workloads, reflecting diverse experimental contexts.

\textbf{Fragmented coverage and limited comparability:} Most studies focus on specific attack types, model classes, or deployment scenarios. Direct comparison between methods is complicated by differing metrics, model sizes, and evaluation environments. While this demonstrates the growing interest in threats to resource consumption, it also highlights the need for consistent, standardized assessment.

These observations motivate the development of a unified evaluation framework that provides a common basis for comparing attacks and defenses and supports reproducible, comprehensive research.

\subsection{Proposed Evaluation Guidelines}

Given the heterogeneity of existing studies, we recommend evaluating resource consumption threats from three complementary perspectives: \emph{model-level behavior}, \emph{hardware-level pressure}, and \emph{application-level impact}. This design follows the intuition that excessive generation first manifests as abnormal decoding behavior at the model level, then translates into measurable computational burden on the serving hardware, and finally affects the quality of service in real deployments. Evaluating all three levels together provides a more complete view of both attack severity and defense effectiveness.

At the \emph{model level}, we recommend reporting \textbf{output length} and \textbf{inference latency}. Output length directly captures the extent to which generation is induced by the attack or controlled by the defense. This metric is particularly important because many resource consumption threats operate by triggering unnecessarily long outputs, repetitive reasoning traces, or non-terminating decoding patterns. It is also highly portable across model families, including standard LLMs, reasoning models, multimodal systems, and agentic components that generate intermediate textual outputs. However, output length alone is insufficient, since the same number of tokens may incur very different computational costs depending on the model architecture, decoding strategy, and modality. We therefore additionally recommend inference latency as a direct measure of time overhead. Latency reflects how long the system takes to process a query under attack or defense, and is especially useful for capturing practical degradation in responsiveness. Together, output length and latency characterize the immediate behavioral footprint of resource amplification at the model side.

At the \emph{hardware level}, we recommend reporting \textbf{GPU utilization} and \textbf{memory utilization}. These metrics capture whether excessive generation is actually converted into pressure on the underlying compute infrastructure. GPU utilization measures how intensively the accelerator is occupied during inference, and is useful for identifying whether an attack induces sustained computational saturation rather than merely producing longer outputs. Memory utilization, including both allocated and peak memory usage where possible, reflects the storage burden created by long contexts, repeated decoding steps, or multimodal feature processing. This metric is especially important for large models and multimodal systems, where memory bottlenecks may become the dominant constraint even before compute is fully saturated. Compared with model-level metrics, hardware-level metrics are more deployment-sensitive, since they depend on serving frameworks, batching policies, quantization, and device types. Nevertheless, they are essential when the goal is to assess the actual infrastructure impact of a threat or the practical efficiency gain of a defense.

At the \emph{application level}, we recommend reporting \textbf{throughput}. Throughput reflects the number of requests or tasks that can be completed within a unit of time, and therefore serves as a direct indicator of service capacity under attack or defense. This metric is particularly important because the ultimate consequence of resource consumption threats is often not limited to a single query becoming slower, but rather to the whole system handling fewer users, fewer concurrent tasks, or fewer agent executions. Throughput thus captures the service-level manifestation of resource abuse, making it especially relevant for online APIs, multi-user serving platforms, and agentic systems with concurrent workflows. In practice, throughput should be interpreted jointly with latency, since a system may maintain acceptable latency for individual queries while still suffering degraded overall capacity under increased workload.

Taken together, these metrics form a layered evaluation framework. Output length and inference latency capture how the model behaves under resource amplification; GPU and memory utilization measure whether this behavior imposes real hardware burden; and throughput reflects whether such a burden ultimately degrades service capacity in practical deployments. This layered design also improves comparability across settings. Model-level metrics are broadly applicable and should be reported in nearly all studies. Hardware-level metrics are particularly important for systems and deployment-oriented evaluations. Application-level metrics are most informative in realistic serving scenarios, especially for production APIs and agentic environments. We therefore recommend that future studies report at least one metric from each level whenever possible, so that resource consumption threats can be assessed not only as generation anomalies, but also as infrastructure and service risks.
\section{Detailed Technical Analysis of Resource Consumption Mechanisms}
\label{sec:appen_mechanisms}
Rather than reiterating the high-level taxonomy presented in the main text, this section provides a \textbf{more fine-grained technical account} of the mechanisms underlying resource consumption behaviors in autoregressive generation. We organize the discussion across four complementary levels of analysis—Markov chain foundations, circuit-level mechanisms, behavioral dynamics, and training-level causes—so that researchers can more precisely trace how repetition and output prolongation arise from theoretical properties, internal model components, emergent generation patterns, and learning objectives, respectively.

\subsection{Markov Chain}
By modeling autoregressive generation as a Markov chain and deriving theoretical upper bounds on the Average Repetition Probability (ARP), ~\cite{fu2021theoretical} identifies \textit{High Inflow Words}—tokens toward which disproportionately many others transition with high probability. These tokens form absorbing-like loops within the transition matrix: once the generative process enters such high-inflow states, it becomes trapped in repetitive cycles rather than progressing toward the EOS token. The proposed rebalanced encoding merges high inflow pairs into single tokens, effectively reducing the inflow term in the ARP bound and significantly lowering repetition rates in both translation and language modeling tasks. From a complementary angle, ~\cite{ildiz2024self} establishes a formal equivalence between single-layer self-attention and Context-Conditioned Markov Chains (CCMC), and analyzes the autoregressive trajectory from a single prompt. Due to the non-mixing nature of CCMC, majority tokens undergo self-reinforcement across generation steps, causing the output distribution to collapse into a singleton or highly limited token subset—providing a principled mathematical account of why LLMs tend to generate repetitive text during prolonged decoding.

\subsection{Circuit-Level Mechanisms}
\paragraph{Repetition Neurons and Features}
Specific neurons within MLP blocks have been identified as direct executors of repetitive content. Differential activation analysis~\cite{hiraoka2025repetition} revealed ``repetition neurons'' distributed throughout the model: intermediate-layer neurons tend to detect repeating patterns, while top-layer neurons drive the model to replicate previous context. Layer-wise causal analysis~\cite{doan2025understanding} further shows that these neurons often function as downstream components of Induction Heads, serving to amplify copying signals. Moving beyond individual neurons, ~\cite{yao-etal-2025-understanding} utilized Sparse Autoencoders (SAEs) to identify latent directions that specifically encode ``repetition features,'' indicating that the model explicitly represents repetitive behaviors within its feature space.
\paragraph{Induction Heads} Through prefix matching score analysis and targeted attention knockouts, ~\cite{crosbie2025induction} provides quantitative evidence that induction heads are the fundamental operators of few-shot in-context learning (ICL). Because output repetition functions as an uncontrolled, degenerate form of ICL, Induction Head Descaling~\cite{wang2025induction} formalized the ``Induction Head Toxicity'' theory to explain this dynamic. From a mechanistic perspective, toxicity occurs when induction heads disproportionately dominate the output logits, thereby suppressing contributions from other attention heads. This logit dominance enforces rigid pattern replication and triggers a rapid entropy collapse in the next-token probability distribution. Adding structural nuance to this, ~\cite{mahaut2025repetitions} revealed that repetition is actually sustained by two distinct, parallel mechanisms. The first is the aforementioned ICL-induced repetition, which relies on a specialized, late-developing circuit of attention heads and MLPs that operate with high prediction confidence. The second is a naturally occurring repetition mechanism that emerges very early in training; rather than using a dedicated circuit, it functions as a degenerate fallback, with attention weights anomalously collapsing onto low-information, structural tokens (e.g., newlines), sustaining repetitive loops even without explicit contextual prompts.

\subsection{Behavioral Dynamics}
By manually constructing pseudo-repetitive data and comparing token probabilities across increasing repetition counts, Pseudo-Repetition Penalization~\cite{xu2022learning} reveals a critical \textit{self-reinforcement effect}: once a model generates one repeated sentence, the probability of continuing to repeat rises almost monotonically with the number of historical repetitions, eventually stabilizing at a high ceiling value. Sentences with higher initial probabilities exhibit stronger self-reinforcement, explaining why maximization-based decoding is particularly prone to sentence-level loops. Based on this finding, the proposed DITTO method trains models to exponentially decay repetition probability on pseudo data, significantly reducing repetitions without sacrificing perplexity. From a complementary angle, uncertainty-driven fallback modeling~\cite{ivgi2024loops} systematically varies model size, pretraining tokens, and instruction tuning to analyze fallback behaviors under \textit{epistemic uncertainty}. Their experiments reveal a consistent ordering: as uncertainty increases during generation, models shift from producing correct facts to hallucinations, then to degenerate text, and finally to verbatim sequence repetitions—positioning repetition as the simplest fallback state when parametric knowledge is exhausted.

\subsection{Training-Level Causes}
Unlikelihood training introduces auxiliary losses to penalize repeated tokens~\cite{welleck2019neural}. Experimental results demonstrate that the approach significantly reduces repetition and dullness while maintaining competitive perplexity and token accuracy, and produces higher-quality generations under standard decoding strategies such as greedy search and beam search. However, such losses alone do not resolve the skewed token-level probabilities inherent in MLE—motivating direct gradient manipulation as an alternative~\cite{lin2021straight}. ScaleGrad~\cite{lin2021straight} directly modifies the gradients of the training objective to encourage the model to assign higher importance to novel tokens during learning. The approach effectively reduces repetition and enhances diversity while maintaining strong performance, as evidenced by both automatic metrics and human evaluations. A more targeted approach, HAPAX~\cite{sahin2025context}, omits the loss contribution of any token predictable by induction heads—effectively ensuring that repeated n-grams within a context window never produce gradient signals. Despite a 66\% reduction in verbatim copying, HAPAX models surpass the vanilla baseline on 13 of 21 abstractive in-context learning tasks, demonstrating that suppressing inductive copying does not compromise broader ICL capabilities. At the data level, Repetition Dropout~\cite{li2023repetition} randomly drops attention to repetitive words during training, directly reducing the model's exposure to repeated patterns. This simple strategy substantially lowers the repetition rate in generated text, and further analysis shows that it provides a unified explanation for prior methods—penalizing training-data repetitions emerges as the common and fundamental factor underlying the effectiveness of high-inflow word mitigation, likelihood objective modifications, and self-reinforcement suppression. Notably, this effect persists across larger model scales and instruction-tuned settings.
\section{Detailed Discussion of Mitigation Methods}
\label{sec:appendix_defense} 

Mitigation is a necessary counterpart to the study of resource consumption threats, especially as recent attacks have shown that resource abuse can escalate from efficiency degradation to severe operational and economic harm. 
The main text has already provided a high-level overview of the current defense landscape, focusing on the available mitigation strategies, their coverage, and the major gaps that remain.
Rather than reiterating the high-level taxonomy in the main text, this appendix provides a \textbf{more fine-grained technical account} of representative defense methods, so that future researchers and practitioners can more easily identify the most suitable interventions for different threat regimes and deployment settings.
Specifically, we expand the discussion along the same two primary threat classes identified earlier---\textit{Overthinking} and \textit{Unbounded Drift}---and further organize existing methods by the system layer at which intervention occurs, including training, decoding, and external control, with emphasis on their \textbf{technical mechanisms}, \textbf{reported effects}, and practical limitations.

\subsection{Defenses against Overthinking}
\paragraph{Large Language Models (LLMs).}
This line of work mitigates overthinking primarily through \emph{prompt-level output control}, where conciseness constraints are explicitly injected into the input to suppress unnecessary reasoning expansion. Constrained-CoT (CCoT)~\cite{nayab2025concise} implements this idea by introducing user-specified length budgets (e.g., 15, 30, or 45 words) directly into the prompt, forcing the model to trade off answer correctness against output brevity during decoding. To evaluate this trade-off, it defines three correctness metrics: Hard-$k$ Concise Accuracy (HCA), which counts only correct responses within a fixed length threshold; Soft-$k$ Concise Accuracy (SCA), which applies an exponential penalty to moderate length violations; and Consistent Concise Accuracy (CCA), which further measures the stability of concise reasoning across generations. It also introduces the Redundancy Mean Score (RMS) and the Information Flow Score to quantify syntactic redundancy and semantic continuity in generated reasoning traces. In essence, these methods do not modify model parameters or decoding dynamics, but instead rely on external prompt-side constraints to compress generation, which makes them lightweight yet inherently limited against adversarially induced length amplification.

\paragraph{Reasoning Large Language Models (RLLMs).}
In reasoning models, a more direct line of defense is to \emph{internalize length control during training}, so that budget awareness becomes an intrinsic capability of the model rather than an external prompt-side constraint. LAPO~\cite{wu2025lapo} implements this idea with a two-stage reinforcement learning procedure. In the first stage, the model performs GRPO rollouts and records the output lengths of correct responses only; the 30th and 70th percentiles are then used to define a filtered target interval $[L_{\min}, L_{\max}]$, and a linear decay reward penalizes generations that fall outside this range. The median feasible length $L_{\text{median}}$ is then carried into the second stage as an explicit self-budgeting target embedded in the prompt, while a Gaussian-style length-adherence reward encourages the model to match its declared budget during generation. This effectively converts length control from post hoc truncation to learned planning behavior, yielding up to 40.9\% token reduction and a 2.3\% accuracy gain on mathematical reasoning benchmarks.

CoT-Valve~\cite{ma-etal-2025-cot} addresses the same problem through \emph{parameter-space controllability} rather than reward shaping. Using LoRA, it identifies a controllable direction in parameter space that governs reasoning length, allowing a single model to compress or expand its chain of thought by adjusting the intervention magnitude. To support this control, the method constructs the MixChain dataset, where each question is paired with reasoning traces of different lengths. Under this design, reasoning compression is achieved as a continuous model-side capability rather than a fixed decoding heuristic: on QwQ-32B, the method reduces GSM8K reasoning length from 741 to 225 tokens with only a 0.15\% accuracy drop, and compresses AIME traces from 6{,}827 to 4{,}629 tokens with only one additional error.

At the decoding stage, existing defenses mitigate overthinking through \emph{lightweight inference-time control} without updating model parameters. A first line of work suppresses explicit reflection triggers once continued reasoning becomes unnecessary. CGRS~\cite{huang2025cgrs}, for example, inserts a probe at structural delimiters and estimates current confidence from output entropy; when the confidence exceeds 0.9, it directly downweights reflection-leading words such as ``Wait'', ``But'', and ``Alternatively'' by assigning them large negative logits, thereby preventing further reflection loops at decoding time.

A second line of work detects and interrupts semantically empty repetition from internal representations. WSC~\cite{xie-etal-2025-word} targets useless self-repetitions that consume decoding budget without adding semantic value. It trains a model-specific linear classifier on hidden states at end-of-chunk positions and triggers chopping when either 2 consecutive long repetitive chunks ($\geq$10 tokens) or 5 short ones ($<$10 tokens) are detected; after interruption, a rescue regeneration prompt such as ``Let me reconsider\ldots'' is appended under a fixed token budget. The self-affirmation suppression approach~\cite{liu2025selfaffirmation} focuses on a narrower redundancy mode, namely reflective steps that merely reaffirm earlier correct content, and suppresses them by exploiting probability biases in their leading words, achieving length reductions of 18.7\% in the train-free setting and 50.2\% in the train-based setting.

A third line of work performs \emph{representation-level steering} of reasoning depth. Controlling Thinking Speed~\cite{lin2025controlling} extracts a steering vector by applying PCA to hidden-state difference vectors between fast- and slow-thinking trajectories, and then adjusts reasoning speed through a strength parameter $\alpha$ ($\alpha>0$ accelerates thinking and $\alpha<0$ slows it down). It further introduces an adaptive variant that estimates real-time reasoning difficulty from the Jensen--Shannon divergence between early-layer and final-layer logits, enabling dynamic switching within a single inference pass and yielding an average $+$1.3\% accuracy gain together with a $-$8.6\% token reduction. FR-Ponder~\cite{he2025frponder} adopts a related latent-control strategy, extracting contrastive steering vectors from step-by-step reasoning versus direct-answer prompts and applying additive hidden-state perturbations through a lightweight controller trained with GRPO, with curriculum learning used to align compute allocation to task difficulty; under this design, it reports 30--40\% token reduction with up to 10\% accuracy improvement on GSM8K, MATH500, and GPQA.

A fourth line of work controls overthinking by allocating \emph{sampling-time budgets}. DSC~\cite{wang-etal-2025-make} first performs batch-level difficulty ranking using the model itself: queries with zero entropy across pre-samples are assigned to an ``Easy'' group and receive only a single chain-of-thought sample, whereas ``Hard'' queries are assigned an initial sampling budget estimated from similarly difficult prior queries. If consensus is not reached, the sampling window is further expanded using a Dirichlet-based stopping rule. This makes DSC fundamentally a batch-level budget scheduler rather than a single-query real-time controller, which limits its applicability in latency-sensitive inference settings.

From an external control perspective, existing methods mitigate overthinking by imposing \emph{budget signals or supervisory constraints outside the model itself}, without directly modifying the underlying reasoning dynamics. TALE~\cite{han-etal-2025-token} implements this idea through token-budget control conditioned on problem complexity: TALE-EP estimates an appropriate budget via zero-shot prompting, while TALE-PT internalizes budget awareness through SFT/DPO post-training, so that the budget can be explicitly injected into the reasoning process as an external control variable. In a simpler form, CCoT~\cite{nayab2025concise} applies the same prompt-side control principle by directly embedding length constraints into user instructions, thereby capping output verbosity without changing model parameters or decoding rules.

A more intervention-oriented line of work introduces \emph{external supervisors} that monitor and terminate redundant reasoning online. MoT~\cite{zhu2025unthinking} follows this design by exploiting the manipulability of special delimiter tokens, the same external control surface used in adversarial BoT-style attacks, to insert a plug-and-play monitoring mechanism over the reasoning process. Rather than compressing reasoning through static prompting alone, it dynamically halts redundant or risky reasoning paths during generation, serving both as an efficiency controller to reduce overthinking and as a safety-oriented supervisor to terminate unsafe reasoning, with a monitoring interval of every 200 tokens.

\paragraph{Multimodal Large Language Models (MLLMs) and Memory Systems.}
In multimodal and long-context settings, overthinking mitigation must address not only unnecessary reasoning depth, but also the growth of retrieved or cached context. A first line of work controls \emph{whether long-form reasoning is invoked at all}. CAR~\cite{lu2025car} implements this through perplexity-based routing, dynamically switching between short responses and chain-of-thought reasoning according to the model's estimated confidence, so that expensive long-form reasoning is triggered only for uncertain inputs.

A second line of work controls \emph{how much contextual state is carried into generation}. NMRet~\cite{bhatnmret} manages long-context overhead through a structured memory architecture consisting of a stateful neural memory for abstract long-term storage, a vector store for contextual retrieval, and a reasoning compressor for intermediate context management, thereby reducing context-window growth during multi-step reasoning. MeVe~\cite{ottem2025meve} approaches the same problem from the retrieval side by inserting explicit verification and compression before generation: it performs initial kNN retrieval, cross-encoder-based relevance filtering, BM25 fallback retrieval, relevance-based context prioritization with redundancy removal, and finally token-budgeted greedy packing to control the final context size. Under this design, irrelevant or low-value context is filtered before entering the model, yielding a 57\% reduction in context token consumption on Wikipedia datasets and 75\% on HotpotQA.

\subsection{Defenses against Unbounded Drift}

The following methods target non-convergent generation, where repetition collapse, infinite loops, or runaway execution amplify resource consumption beyond output inefficiency and into crash-level system pressure.

\paragraph{Large Language Models (LLMs).}
For non-convergent generation in general LLMs, existing defenses mainly intervene at the training and decoding stages. At training time, the core idea is to reshape the learning objective so that repetitive patterns are explicitly suppressed during optimization. Unlikelihood training~\cite{welleck2019neural} implements this by introducing negative updates at two granularities. At the token level, previously generated tokens in the context are treated as negative candidates, and an unlikelihood term is added to the standard MLE objective to penalize high probability assigned to those historical tokens, thereby suppressing local repetition and high-frequency token dominance. At the sequence level, tokens belonging to repeated $n$-grams in the decoded outputs are marked as negative samples during fine-tuning, extending this principle to longer structural repetition patterns. These two objectives are combined in training, though they exhibit different limitations: token-level optimization suffers from a distribution mismatch between training and generation, whereas sequence-level optimization requires decoding full sequences within the training loop and is therefore substantially more expensive. Under beam search, the combined design reduces the 4-gram repetition rate from 0.442 to 0.013 and increases the number of unique generated tokens by 77\%.

At decoding time, the focus shifts from suppressing repetition in the learned distribution to controlling it online during generation. RAP~\cite{huang2025rap} formalizes this process through systematic tuning of the Repetition Penalty Parameter (RPP). Its core measurement component, ReDA, computes a repetition ratio (RR) for each output sequence by detecting not only standard textual repetition but also consecutive non-word-character repetition and space-free long-form repetition through regular-expression-based matching. Given RR, RAP selects the penalty strength by maximizing the score $P \times F(\mathrm{RR})$, where $P$ denotes task performance and $F(\mathrm{RR})$ is a penalty function over repetition severity. Among five candidate forms---linear, quadratic, cubic, logarithmic, and exponential---the cubic function $(1-\mathrm{RR})^3$ yields the best trade-off according to the reported ablation results.

For more severe resource exhaustion and DoS-style threats, the most mature defenses operate at the \emph{system level}, where intervention is applied during serving rather than through model retraining. A representative direction is \emph{online loop detection}. RecurrentDetector~\cite{yu2025breaking} monitors Transformer activation states at each generation step with a lightweight MLP classifier and terminates generation once the cosine similarity between the current and previous states exceeds 0.95, treating high state recurrence as a signal of non-convergent looping. Under this design, it achieves 95.24\% detection accuracy with a 2.59\% false positive rate and only 0.36ms additional latency. Its scope, however, is limited by strong white-box assumptions: it cannot be applied to closed-source APIs such as GPT-4, may be bypassed by polymorphic attacks that preserve semantic loop structure while varying surface tokens, and has been evaluated on only 6 open-source architectures across 4,000 prompts.

A complementary direction performs \emph{resource-aware serving control} under adversarial load. PD\textsuperscript{3}F~\cite{zhang2025pd3f} implements this through a two-stage pipeline. On the input side, it computes a Resource Index from five features — total inference time, peak GPU memory, peak GPU utilization, input length, and output length — and uses this signal to guide dynamic request polling under high-concurrency, malicious prompts. On the output side, it applies an Adaptive End-Based Suppression mechanism that explicitly amplifies the EOS logit to terminate maliciously prolonged generation. Under AutoDoS-style attacks, this design improves legitimate-user throughput by up to 500\%, achieves attack-detection accuracy above 99\% across AutoDoS, GCG-DoS, and P-DoS settings, and has been validated across six open-source LLMs. Its main limitation is threshold sensitivity: the IQR-based decision rule may misclassify legitimate but computationally intensive requests, such as long code generation, leading to substantial false positives in real deployments.

At a broader infrastructure level, queue-based web service architectures~\cite{shahriar2025vulnerability} mitigate overload by \emph{decoupling request admission from direct model execution}. Instead of treating every request as an immediately executable generation job, they use distributed queuing to smooth bursty demand, isolate overload pressure from the serving backend, and preserve stable, near-linear scalability under extreme workloads. Compared with model-centric defenses, this line of work does not directly suppress malicious generation behavior but provides an engineering-level containment mechanism for maintaining service stability when resource abuse cannot be fully blocked upstream.

\paragraph{Reasoning Large Language Models (LRMs).}
Purpose-built defenses against crash-level and non-terminating behavior in reasoning models remain largely absent. Existing mitigation methods only provide \emph{incidental coverage} by suppressing semantically empty self-repetition during overthinking control. For example, the self-affirmation suppression approach~\cite{liu2025selfaffirmation} and WSC~\cite{xie-etal-2025-word} can interrupt mild repetitive loops because both target redundant reflective steps that consume decoding budget without adding semantic value. However, these methods are not designed to handle adversarially induced collapse. WSC relies on a model-specific classifier, and its regeneration stage may itself enter a loop, while CGRS depends on hardcoded reflection triggers and can therefore be bypassed when prompt injection suppresses or substitutes those exact tokens. As a result, current reasoning-model defenses do not yet provide reliable protection against sustained crash-style attacks, leaving a clear gap between overthinking mitigation and true non-termination defense.

\paragraph{Multimodal Large Language Models (MLLMs).}
For latency-oriented resource threats in multimodal and edge-deployed systems, existing defenses mainly intervene through \emph{hardware-aware adversarial training}. Background-attentive adversarial training~\cite{wang2025cantslowmedown} implements this idea by explicitly coupling perturbation robustness with device-level capacity constraints across heterogeneous GPUs. Technically, it uses binary masks to concentrate perturbation awareness on vulnerable background regions, where latency attacks often induce excessive post-processing overhead, and incorporates objectness loss as an auxiliary signal to distinguish true objects from attacker-induced phantom detections. Under this design, the defense improves both robustness and serving efficiency: on Jetson Orin NX, it restores processing speed from 13~FPS to 43~FPS, achieves 8--10\% higher robust accuracy than MTD and OOD, and incurs only 4.4\% clean accuracy loss. The reported evaluation covers multiple YOLO-based detectors, including YOLOv3, YOLOv5, and YOLOv8, across autonomous driving and general object detection settings.
\section{Benchmarks and Datasets}
\label{sec:appen_bench}

Existing benchmarks for resource consumption in large language models remain relatively limited. A small number of benchmarks explicitly focus on excessive generation and DoS–style resource consumption. For example, BenchOverflow~\cite{feiglin2026benchoverflow} measures the overflow phenomenon, in which benign plaintext prompts trigger abnormally long outputs. Prompt-Induced Over-Generation as Denial-of-Service~\cite{guo2025prompt} constructs an attack-side benchmark that evaluates how prompts can induce over-generation under black-box access. 

In contrast, a larger body of benchmarks focuses on efficiency issues arising from long-chain reasoning. Stop Overthinking~\cite{sui2025stop} surveys efficient reasoning methods and highlights the prevalence of excessive reasoning behavior in chain-of-thought generation. EffiReason-Bench~\cite{huang2025effireason} proposes a unified benchmark for evaluating efficiency–performance trade-offs across multiple reasoning tasks and models. SafeChain~\cite{jiang2025safechain} studies the safety implications of long chain-of-thought reasoning and introduces datasets for evaluating safety and robustness in reasoning-intensive settings.

Overall, while recent work has begun to establish evaluation frameworks for both resource-consumption attacks and reasoning efficiency, existing benchmarks remain fragmented across different research objectives. A unified evaluation paradigm for resource-aware generation and robustness is still largely lacking.

\section{AI Writing Assistance Disclosure}
We used AI tools solely for language polishing to improve clarity and readability. The AI tools did not contribute to the scientific content, ideas, analyses, or conclusions of this work.

\begin{table*}[t]
\centering
\resizebox{\textwidth}{!}{%
\begin{tabular}{@{}p{3cm}|p{2cm}|p{4cm}|p{4cm}|c@{}}
\toprule
\textbf{Method} & \textbf{Dataset} & \textbf{Evaluation Metrics} & \textbf{Models} & \textbf{Efficiency} \\ \midrule
\rowcolor[HTML]{EFEFEF} 
{ \textbf{A sloth (Multi-exit) \cite{hong2020panda}}} & { CIFAR-10, ImageNet, Tiny-ImageNet} & { Exit Index, Avg Latency, Energy} & { MSDNet, ResNet (Multi-exit)} & { N/A} \\
{ \textbf{SkipSponge \cite{lintelo2024skipsponge}}} & { CIFAR-10, SVHN} & { FLOPs Inflation, Accuracy} & { ResNet, WideResNet} & { N/A} \\
\rowcolor[HTML]{EFEFEF} 
{ \textbf{On-Device Sponge \cite{wang2023energy}}} & { Speech Commands, IMDB} & { Battery Drain, Execution Time} & { MobileNetV2, ShuffleNet} & { N/A} \\
{ \textbf{Dynamic Routing \cite{chen2023dark}}} & { CIFAR-100, ImageNet} & { Avg Layers, Activation Ratio} & { GaterNet, SkipNet} & { N/A} \\
\rowcolor[HTML]{EFEFEF} 
{ \textbf{Sponge Poisoning \cite{wang2023energy}}} & { ImageNet, CIFAR} & { Latency, GPU Energy} & { VGG, ResNet} & { N/A} \\
{ \textbf{Energy Backdoor \cite{meftah2025energy}}} & { GLUE, SQuAD} & { Energy Increase, Clean Accuracy} & { BERT, RoBERTa} & { 512*} \\
\rowcolor[HTML]{EFEFEF} 
{ \textbf{Sensing AI Sponge \cite{hasan2025sponge}}} & { Audio/Sensor Data} & { Sensor Power, Pruning Ratio} & { CNN, DeepSense} & { N/A} \\
{ \textbf{Transslowdown \cite{chen2021transslowdown}}} & { WMT'14 (En-De), WMT'16 (En-Ro)} & { Translation Latency, Token Count, BLEU Score} & { Transformer, LSTM-based NMT} & { 1024} \\
\rowcolor[HTML]{EFEFEF} 
{ \textbf{NMTSloth \cite{chen2022nmtsloth}}} & { WMT'17 (En-De), WMT'19 (Zh-En)} & { Real-world Latency, Energy, Response Time} & { Fairseq, OpenNMT, MarianNMT} & { \textbf{80x}} \\ \bottomrule
\end{tabular}%
}
\caption{Summary of representative resource consumption attacks in early adaptive architectures.}
\label{tab:attack_summary_early}
\end{table*}
\begin{table*}[t]
\centering
\resizebox{\textwidth}{!}{%
\begin{tabular}{@{}p{3cm}|p{2cm}|p{4cm}|p{4cm}|c@{}}
\toprule
\textbf{Method} & \textbf{Dataset} & \textbf{Evaluation Metrics} & \textbf{Models} & \textbf{Efficiency} \\ \midrule
\rowcolor[HTML]{EFEFEF} 
\textbf{AutoDoS \cite{zhang2025crabs}} & Chatdoctor, MMLU, Hellaswag, Codexglue, GSM & Attack Success Rate, Safety Compliance Rate, Task Success, Token Usage, Latency, Compute Cost & GPT, Llama, Qwen, Deepseek, Ministral, Gamma & 8192 \\
\textbf{Repeated Token \cite{yona2025interpreting}} & OpenWebText, Custom Repetition Dataset & Attention Scores, Loss, Output Length & Llama-3, GPT-2, Pythia & 8192 \\
\rowcolor[HTML]{EFEFEF} 
\textbf{Non-halting Queries \cite{hammouri2025non}} & RAG Systems, 10k Probe Dataset & Avg Token Length, Generation Success & GPT-4o, Llama-3, Gemma-2 & 8192 \\
\textbf{ThinkTrap \cite{li2025thinktrap}} & Sponge, LLMEffiChecker & Throughput, Response Latency, ASR & GPT-4o, Gemini 2.5 Pro, DeepSeek R1 & 8192 \\
\rowcolor[HTML]{EFEFEF} 
\textbf{Crabs (AutoDoS) \cite{zhang2025crabs}} & MMLU, GSM8K, Chatdoctor, Codexglue & Latency Inflation, Output Length ↑, GPU Memory & GPT-4, Llama-3, Qwen-2, DeepSeek, Gamma & 8192 \\
\textbf{BitHydra \cite{yan2025bithydra}} & MMLU, GSM8K & Inference Cost, Bit-flip Rate, ASR & Llama-2, OPT, Bloom & 8192 \\
\rowcolor[HTML]{EFEFEF} 
\textbf{Coercing LLMs \cite{geiping2024coercing}} & AdvBench, Vicuna-bench & ASR, Safety, Efficiency Loss & Vicuna, Llama-2, GPT-3.5 & 4096 \\
\textbf{Engorgio Prompt \cite{dong2024engorgio}} & ShareGPT, Alpaca & Response Length, Energy Draw & Llama-3, Mistral, GPT-4 & 8192 \\
\rowcolor[HTML]{EFEFEF} 
\textbf{LLMEffiChecker \cite{feng2024llmeffichecker}} & Efficiency-Bench, WikiText & Delay, Energy, Throughput & Llama-2, Vicuna, Alpaca & 8192 \\
\textbf{LoopLLM \cite{li2025loopllm}} & MT-Bench, Chatbot Arena & Repetition Ratio, Energy Latency & Llama-3, Qwen-2, Phi-3 & 8192 \\
\rowcolor[HTML]{EFEFEF} 
\textbf{DoS Poisoning \cite{gao2024denial}} & Anthropic HH-RLHF & Inference Latency, Poisoning Ratio & GPT-Neo, RoBERTa & 1024 \\ \bottomrule
\end{tabular}%
}
\caption{Summary of representative resource consumption attacks in LLMs.}
\label{tab:attack_summary_llm}
\end{table*}
\begin{table*}[t]
\centering
\resizebox{\textwidth}{!}{%
\begin{tabular}{@{}p{3cm}|p{2cm}|p{4cm}|p{4cm}|c@{}}
\toprule
\textbf{Method} & \textbf{Dataset} & \textbf{Evaluation Metrics} & \textbf{Models} & \textbf{Efficiency} \\ \midrule
\rowcolor[HTML]{EFEFEF} 
\textbf{Hidden Tail \cite{zhang2025hidden}} & MS-COCO & Attack Success Rate,Output Length, Latency, Visible length, Response Quality & Qwen2.5-VL, MiMo-VL-7B-RL,Gemma-3-4B-IT & 1831 \\
\textbf{NICGSlowDown \cite{chen2022nicgslowdown}} & MS-COCO, Flickr8k & I-Loops, I-Latency (CPU/GPU) & ResNext-LSTM, GoogLeNet-RNN, MobileNets-LSTM & N/A \\
\rowcolor[HTML]{EFEFEF} 
\textbf{Verbose Images \cite{gao2024inducing}} & MS-COCO, ImageNet & Sequence Length, Energy, Latency, Uncertainty & BLIP,BLIP-2, InstructBLIP, MiniGPT-4 & 8x \\
\textbf{Verbose Samples \cite{gao2024energy}} & MSVD, TGIF & Length, Latency, Energy & VideoChat-2,Video-Vicuna,Video-LLaMA & 4x \\
\rowcolor[HTML]{EFEFEF} 
\textbf{Uniform Inputs \cite{muller2024impact}} & ImageNet, & Activation Density, Activation Sparsity & ResNet, DenseNet, MobileNetV2 & N/A \\
\textbf{LingoLoop \cite{fu2025lingoloop}} & MS-COCO, ImageNet & Tokens, Energy Latency & InstructBLIP, Qwen2.5-VL, InternVL3 & 2048 \\
\rowcolor[HTML]{EFEFEF} 
\textbf{Phantom Sponges \cite{shapira2023phantom}} & Berkeley Deep Drive(BDD), Mapillary Traffic Sign Dataset (MTSD), LISA, PASCAL VOC & Number of candidates, Processing Time, Detection Recall & YOLOv5, YOLOv3, YOLOv4 & N/A \\
\textbf{Enhanced PhantomSponges \cite{schoof2024beyond}} & Berkeley Deep-Drive & Number of candidates, Processing Time, Detection Recall & YOLOv5 & N/A \\
\rowcolor[HTML]{EFEFEF} 
\textbf{RECITE \cite{gao2025resource}} & ImageNet, MMLU, HumanEval, GSM8K & Attack Success Rate, Service response latency, GPU utilization, Average generation length & InstructBLIP, LLaVA, Qwen-VL & 2048 \\
\textbf{EO-VLM \cite{seo2025eo}} & - & GPU Power, Inference Time & YOLOv8, MASKDINO, Detectron2 & N/A \\
\rowcolor[HTML]{EFEFEF} 
\textbf{QuantAttack \cite{baras2025quantattack}} & ImageNet & GPU Memory, Processing Time, Energy, Outlier Count, Accuracy & Vision Transformer (ViT), Data-efficient image Transformer (DeiT) & N/A \\
\textbf{VLMInferSlow \cite{wang2025vlminferslow}} & MS-COCO, ImageNet & I-length I-latency I-energy & FLAMINGO, BLIP, GIT, FLORENCE & N/A \\
\rowcolor[HTML]{EFEFEF} 
\textbf{Sponge Examples \cite{shumailov2021sponge}} & CIFAR-10, SVHN, WikiText-103, En-De & Energy Consumption, Latency, Memory Access & VGG-16, ResNet-18, GPT-2 & 512 \\ \bottomrule
\end{tabular}%
}
\caption{Summary of representative resource consumption attacks in MLLMs.}
\label{tab:attack_summary_mllm}
\end{table*}
\begin{table*}[t]
\centering
\resizebox{\textwidth}{!}{%
\begin{tabular}{@{}p{3cm}|p{2cm}|p{4cm}|p{4cm}|c@{}}
\toprule
\textbf{Method} & \textbf{Dataset} & \textbf{Evaluation Metrics} & \textbf{Models} & \textbf{Efficiency} \\ \midrule
\rowcolor[HTML]{EFEFEF} 
\textbf{OverThink \cite{kumar2025overthinkslowdownattacksreasoning}} & FreshQA, SQuAD, MuSR & Reasoning token amplification, Answer correctness, Guardrail evasion, Defense evaluation & o1, o1-mini, DeepSeek-R1 & 46x \\
\textbf{Excessive Reasoning Attack \cite{si2025excessivereasoningattackreasoning}} & GSM8K, ORCA & Reasoning length and output length, utility performance (accuracy) & DeepSeek-R1-Distill-LLaMA, DeepSeek-R1-Distill-Qwen, o1-mini, o3-mini, DeepSeek-R1, QWQ & 9x \\
\rowcolor[HTML]{EFEFEF} 
\textbf{BadReasoner \cite{yi2025badreasonerplantingtunableoverthinking}} & - & Reasoning verbosity, answer correctness, controllability & Marco-o1, QwQ, DeepSeek-R1 & N/A \\
\textbf{BadThink \cite{liu2025badthinktriggeredoverthinkingattacks}} & MATH-500, GSM8K & ASR, RIR, TAC, BAD, SD & DeepSeek-R1-Distill-Qwen (1.5B/7B/14B/32B), OpenR1-Qwen-7B, Light-R1-7B-DS & 63.85x \\
\rowcolor[HTML]{EFEFEF} 
\textbf{ExtendAttack \cite{zhu2025extendattackattackingserverslrms}} & AIME 2024, AIME 2025, HumanEval, BigCodeBench-Complete & Response Length, Latency, Accuracy(Pass@1) & o3, o3-mini, QwQ-32B, Qwen3-32B & N/A \\
\textbf{RepetitionCurse \cite{huang2025repetitioncursemeasuringunderstandingrouter}} & - & prefill latency,TTFT,router imbalance & Mixtral-8x7B series, Qwen3-30B-A3B series, GPT-OSS-20B/120B, Kimi-Linear-Instruct, DeepSeek-V2-Lite, Llama-4-Scout-17B-16E-Instruct & N/A \\ \bottomrule
\end{tabular}%
}
\caption{Summary of representative resource consumption attacks in RLLMs.}
\label{tab:attack_summary_rllm}
\end{table*}
\begin{table*}[t]
\centering
\resizebox{\textwidth}{!}{%
\begin{tabular}{@{}p{3cm}|p{2cm}|p{4cm}|p{4cm}|c@{}}
\toprule
\textbf{Method} & \textbf{Dataset} & \textbf{Evaluation Metrics} & \textbf{Models} & \textbf{Efficiency} \\ \midrule
\rowcolor[HTML]{EFEFEF} 
\textbf{SlowLiDAR \cite{Liu_2023_CVPR}} & KITTI, nuScenes & Runtime latency, Imperceptibility & PointPillars, SECOND, PV-RCNN & 2.7x \\
\textbf{CORBA \cite{zhou2025corbacontagiousrecursiveblocking}} & LLM-MAS simulation tasks & P-ASR, PTN, Availability degradation & GPT-4o-mini, GPT-4, GPT-3.5-turbo, Gemini-2.0-Flash, Qwen2.5-14B-Instruct, Llama-3.1-70B-Instruct, Gemma-2-27B-it & {\color[HTML]{1F1F1F} N/A} \\
\rowcolor[HTML]{EFEFEF} 
\textbf{CP-FREEZER \cite{wang2025cpfreezerlatencyattacksvehicular}} & OPV2V & End-to-end latency, Attack success rate, Frame processing time & OpenCOOD & 90x \\
\textbf{SlowTrack \cite{Ma_Wang_Chen_Shen_2024}} & MOT17 & Latency increase (R-Lat), Imperceptibility, System-level crash rate & SORT, FairMOT, ByteTrack, BoT-SORTSORT (YOLOv5 detector), FairMOT, ByteTrack, BoT-SORT & {\color[HTML]{333333} 4x} \\
\rowcolor[HTML]{EFEFEF} 
\textbf{LeechHijack \cite{zhang2025leechhijackcovertcomputationalresource}} & GAIA, GPQA, MMLU & Attack success rate, Resource overhead, Detectability (stealthiness) & DeepSeek, Qwen, GPT, Gemini & {\color[HTML]{1F1F1F} N/A} \\
\textbf{Overload \cite{Chen_2024_CVPR}} & MS COCO dataset & Inference time, NMS latency & YOLOv5 & 10x \\
\rowcolor[HTML]{EFEFEF} 
\textbf{Clawdrain \cite{dong2026clawdrainexploitingtoolcallingchains}} & Real OpenClaw agent workloads & Token amplification, Cost overhead, Attack success rate, Stealthiness & OpenClaw v2026.2.9 & $\sim$9x \\ \bottomrule
\end{tabular}%
}
\caption{Summary of representative resource consumption attacks in Agents.}
\label{tab:attack_summary_agent}
\end{table*}
\begin{table*}[t]
\centering
\resizebox{\textwidth}{!}{%
\begin{tabular}{@{}p{2cm}|p{3cm}|p{3cm}|p{3cm}|p{3cm}|c@{}}
\toprule
\textbf{Method} & \textbf{Dataset} & \textbf{Evaluation Metrics} & \textbf{Models} & \textbf{Efficiency} & \textbf{Model Class} \\ \midrule
\rowcolor[HTML]{EFEFEF} 
\textbf{PD$^3$F \cite{zhang2025pd3f}} & MMLU, Hellaswag, HumanEval, GSM, GPQA & Attack Success Rate, Task Success, Token Usage, Latency & Llama, Mistral, Qwen & Total Time 50\% $\downarrow$ & LLM \\
\textbf{RecurrentD- etector \cite{yu2025breaking}} & Custom trigger dataset (2388 inputs), ShareGPT & Accuracy, F1, FPR, Recall, trigger attempts, latency & Llama-3/2 (7/13B), Vicuna-v1.5 (7/13B), Gemma-2 (2B), GPT-4o/mini & RecurrentGenerator: avg attempts 272.1 vs 1679.1 random; RecurrentDetector: 0.36 ms inference & LLM \\
\rowcolor[HTML]{EFEFEF} 
\textbf{CCoT \cite{nayab2025concise}} & GSM8K, SVAMP, ASDIV & Accuracy, gen time, token count, HCA/SCA/CCA, RMS, Info Flow & Llama2-70b/7b, Falcon-40b/7b, Vicuna-13b & 5.12 s generate time $\downarrow$, 4.41\% ACC $\uparrow$ & LLM \\
\textbf{CoT-Valve \cite{ma-etal-2025-cot}} & GSM8K, PRM800K (ground truth), MixChain C/Z; Eval: GSM8K, AIME24 & Pass@1, token count, ACU & QwQ 32B Preview, DeepSeek R1 Distill Llama 8B, LLaMA 3.1/3.2 (8B/1B), Qwen2.5 32B (w/ LIMO) & Tokens usage 69.6\%$\downarrow$ & LLM \\
\rowcolor[HTML]{EFEFEF} 
\textbf{FR-Ponder \cite{he2025frponder}} & GSM8K, MATH500, GPQA & Acc, avg tokens, avg FLOPs (log) & LLaMA 3 (8/70B), Qwen 2.5 (0.5/3/7B) & 30–50\% token $\downarrow$ & LLM \\
\textbf{DSC \cite{wang-etal-2025-make}} & MATH, GSM8K, CSQA, SQA, Last Letter, Coin Flip & Acc, cost (\$), tokens, time & GPT 3.5 Turbo, GPT 4, Mistral 7B Instruct v0.3 & Cost $\downarrow$ 65\% (GPT 4), 56\% (GPT 3.5) & LLM \\
\rowcolor[HTML]{EFEFEF} 
\textbf{MeVe \cite{ottem2025meve}} & English Wikipedia (first 100 articles), HotpotQA subset & Avg context tokens, retrieval time, grounding/relevance proxies & Embedding, cross encoder, tokenizer, BM25 (fallback) & Context tokens $\downarrow$ 57.7\% (Wiki) \& 75\% (HotpotQA) & LLM \\
\textbf{Unlikelihood Training \cite{welleck2019neural}} & Wikitext 103, GPT 2 fine tuning corpus & seq/token repetition metrics, ppl, acc, human win rate & 16 layer Transformer, GPT 2 (medium pre trained) & Token level: 150k updates; seq level: 1.5k updates & LLM \\
\rowcolor[HTML]{EFEFEF} 
\textbf{NMRet \cite{bhatnmret}} & CoQA, ai arxiv2, TextVQA & RAGAS metrics & GeminiLLM, Titans pytorch NeuralMemory, Qdrant/Chroma, CLIPEmbeddings, LightThinkerCompressor & LightThinker Compressor reduces token footprint for KV cache & LLM \\ \bottomrule
\end{tabular}%
}
\caption{Summary of representative defenses against resource consumption threats (Part 1).}
\label{tab:deffence_summary1}
\end{table*}
\begin{table*}[t]
\centering
\resizebox{\textwidth}{!}{%
\begin{tabular}{@{}p{2cm}|p{3cm}|p{3cm}|p{3cm}|p{3cm}|c@{}}
\toprule
Method & \textbf{Dataset} & \textbf{Evaluation Metrics} & \textbf{Models} & \textbf{Efficiency} & \textbf{Model Class} \\ \midrule
\rowcolor[HTML]{EFEFEF} 
\textbf{RAP \cite{huang2025rap}} & QA: WebQSP, MS MARCO; MT: MAC (Chinese English) & RR, RAP score, F1/BERT F1/COMET & Llama 2 (7/13/70B), Llama 3 (8/70B), Llama3.1 70B, Gemma 1.1 (2/7B), Phi 3/3.5 mini (3.8B), Mistral 7B v0.2/0.3 & RR $\downarrow$ up to 93\% (WebQSP) and 74\% (MS MARCO) & LLM \\
\textbf{Queue-based \cite{shahriar2025vulnerability}} & Queue based Web Service (Sidekiq) with custom load tests (120/300/600 req) & Total Time, Avg Time per Request & GPT 2, BLOOM, OPT & avg time 21–35 sec & LLM \\
\rowcolor[HTML]{EFEFEF} 
\textbf{CAR \cite{lu2025car}} & Multimodal (DocVQA, ChartQA, FUNSD, etc.) \& Text (GSM8K, MathQA, StrategyQA) + pilot tasks & ACC (VQA/KIE), EM, token count, PPL & Qwen2.5 0.5B/7B, Llama3.1 8B, Qwen2 VL 7B & Token $\downarrow$ 21–39\%, ACC $\uparrow$ 5.5–6.9\% & LLM, MLLM \\
\textbf{TALE \cite{han-etal-2025-token}} & GSM8K, GSM8K Zero, MathBench (Arithmetic/Middle/High/College) & ACC, output tokens, expense & GPT 4o mini, Yi lightning, GPT 4o, o3 mini, Llama 3.1 8B Instruct & token $\downarrow$ 64–67\%, expense $\downarrow$ 45–59\% & LLM, RLLM \\
\rowcolor[HTML]{EFEFEF} 
\textbf{Underload \cite{wang2025cantslowmedown}} & PASCAL-VOC, COCO, Berkeley DeepDrive (BDD) & mAP50, FPS, NMS latency, compute overhead, memory transfer & YOLOv3, YOLOv5, YOLOv8 & FPS 13→43 on Jetson Orin NX & MLLM \\
\textbf{Thinking Speed \cite{lin2025controlling}} & MATH 500, AIME24/25, GPQA Diamond, LiveCodeBench & Pass@1, token count, latency, mode switch & DeepSeek-R1-Distill-Qwen-7B/32B, QwQ-32B, Qwen3-8B & 1.26\% acc $\uparrow$ and 8.56\% token $\downarrow$ & RLLM \\
\rowcolor[HTML]{EFEFEF} 
\textbf{Self-Affirmation \cite{liu2025selfaffirmation}} & AIME24, AMC23, GSM8K, MATH500, GPQA D (train free \& train based) & Acc, token count, LR ratio & R1 Distill Qwen (1.5B/7B/32B), QwQ 32B, Qwen3 32B & Train free: 8.4–18.7\% token $\downarrow$ & RLLM \\
\textbf{LAPO \cite{wu2025lapo}} & Train: 10k math (6k DeepScaleR, 4k MATH); Eval: MATH500, AIME24, AMC23, OlympiadBench, GPQA & Pass@1, token count, trade off & DeepSeek R1 1.5B, DeepScaleR 1.5B Preview & Token $\downarrow$ up to 40.9\% & RLLM \\
\rowcolor[HTML]{EFEFEF} 
\textbf{MoT \cite{zhu2025unthinking}} & AIME 2024, MATH500, StrongReject, Harmbench, WildJailbreak & ASR, RTC, RPC, C ACC, Min Steps, Refusal, Harmful Score & DeepSeek R1 (1.5/7/14/32B), Light R1 7B DS, Open R1 7B, QwQ 32B, Marco o1 7B & ASR > 90\%, RTC $\sim$ 80\% $\downarrow$, RPC –90\% & RLLM \\
\textbf{Word Salad Chopper \cite{xie-etal-2025-word}} & GSM8K, MATH500, AIME25, GPQA Diamond & ACC, length compression ratio & DeepSeek R1 Distill Qwen (1.5B/7B), DeepSeek R1 Distill Llama 8B, Qwen3 8B & Length compression 13–57\% & RLLM \\ \bottomrule
\end{tabular}%
}
\caption{Summary of representative defenses against resource consumption threats (Part 2).}
\label{tab:deffence_summary2}
\end{table*}

\end{document}